\documentclass[preprint,12pt]{aastex}

\newcommand{\eg}{{\sl e.g.}}
\newcommand{\ie}{{\sl i.e.}}
\newcommand{\etal}{{\rm et al.}}
\newcommand{\chis}{$\chi^{2}~$}
\newcommand{\chir}{${\chi}_{\nu}^{2}$}

\newcommand{\xmm}{{\it XMM-Newton}}
\newcommand{\mrk}{Mrk~766}
\newcommand{\ark}{Ark~564}

\newcommand{\ltsimeq}{\raisebox{-0.6ex}{$\,\stackrel
        {\raisebox{-.2ex}{$\textstyle <$}}{\sim}\,$}}
\newcommand{\gtsimeq}{\raisebox{-0.6ex}{$\,\stackrel
        {\raisebox{-.2ex}{$\textstyle >$}}{\sim}\,$}}

\slugcomment{To appear in the Astrophysical Journal}
\shorttitle{AGN Fourier Resolved Spectra} 
\shortauthors{Papadakis et al.}

\begin{document}

\title{Fourier-Resolved Spectroscopy of AGN using XMM-Newton
data: I. The 3-10 keV band results}

\author{I.E. Papadakis\altaffilmark{1,2}, Z. Ioannou\altaffilmark{2,1},
        D.Kazanas\altaffilmark{3}}
\altaffiltext{1}{Physics Department, University of Crete, Heraklion,
         71003, Crete, Greece}
\altaffiltext{2}{IESL, Foundation for Research and Technology-Hellas, 711 10
        Heraklion,Crete, Greece}
\altaffiltext{3}{Laboratory for High Energy Astrophysics, NASA, Goddard
         Space Flight Center, Code 661, Greenbelt, MD 20771, USA}

\begin{abstract} We present the results from the  Fourier Resolved Spectroscopy
of archival \xmm ~ data of five AGN,  namely, \mrk, NGC 3516, NGC 3783, NGC
4051 and \ark. This work supplements the  earlier study of MCG-6-30-15 as well
as those of several Galactic Black Hole Candidate sources. Our results exhibit
much larger diversity than those of  Galactic sources, a fact we attribute to
the diversity of their masses. When we take into account this effect and
combine our results with those from Cyg X-1, it seems reasonable to conclude
that, at high frequencies, the slope of the Fourier-resolved spectra in
accreting black hole systems decreases with increasing frequency as  $\propto
f^{-0.25}$, irrespective of whether the system is in its High or Low state.
This result implies that the flux variations in AGN are accompanied by complex
spectral slope variations as well. We also find that the Fe K$\alpha$ line in
\mrk, NGC~3783 and NGC~4051 is variable on time scales $\sim 1$ day -- 1 hour.
The iron fluorescence line is absent in the  spectra of the highest frequencies,
and there is an indication that, just like in Cyg X-1, the equivalent width of
the line in the Fourier-resolved of AGN decreases with increasing frequency.
\end{abstract}

\keywords{Galaxies: Active, Galaxies: Seyfert, X--Rays: Galaxies}


\section{Introduction}

Our generic notion of the central engine of Active Galactic Nuclei (AGN) and,
in general, also that of compact Galactic sources such as neutron stars and
black holes powered by accretion, involves a geometrically thin, optically
thick accretion disk that emits locally as a black body ``sandwiched" between
a hot ($\sim 10^9$ K) corona which up-Comptonizes the disk thermal radiation to
produce the ubiquitous X--ray emission associated with this class of objects.
Despite the compelling theoretical arguments in support of this picture,
independent supporting evidence of the specific geometrical arrangement is
hard to come by. For one, only a small fraction of the very broad band
multicolor  black body spectrum of the putative disk is covered by our
instruments making hard the detailed assessment of the form of its spectrum.
Second, the  Comptonization spectrum of the hot corona provides information for
only the  integral of the Comptonization parameter along our line of sight,
rather than the conditions of the local plasma.

Independent support of the above generic picture has been sought in distinct
spectral and timing signatures  implied by the specific geometric arrangement.
More specifically, the  Fe K$\alpha$ fluorescence line at $E = 6.4$ keV as well
as the so-called reflection `hump', the  product of reprocessing the X--rays of
the hot corona by the matter of the underlying accretion disk, are thought to
provide a direct measure of the above geometry. Furthermore, the Keplerian
motion of the reprocessing matter in the disk would lead to a rather broad
profile  for this spectral feature. Indeed, the detection of such a feature in
the ASCA  spectra  of many Seyfert galaxies (\eg\ Nandra \etal\ 1997) seems to
provide a confirmation of these notions. At the same time it has been argued
that  detailed X--ray spectroscopy of the iron line emission features can
provide  information on the location and kinematics of the cold material within
a few gravitational radii of the event horizon.

In addition to the above features, a measure of the size of the X--ray emitting
region, independent of its spectral properties, can be obtained from time
variability studies. In the specific case of emission produced by the
Comptonization process, Kazanas, Hua \& Titarchuk (1997) argued that timing
observations are the only way to estimate the density of  the emitting plasma
(in contrast to its column density provided by the  Comptonization spectra) and
thus peek into the dynamics of the accretion  flow. The principal method   of
characterizing the variability of accretion powered sources has been  the
measurement of the Power Spectral Density function (PSD)  of their X--ray light
curves, which in the case of AGN, were shown to be simple power laws with
occasional ``breaks" in the slope at sufficiently low frequencies. However,
unlike the Comptonization spectra whose slopes and cut-offs can be related
directly to the physical parameters of the  emitting plasma, there is no simple
model that relates in a direct fashion the shape of the PSD to the physics of
the accretion flow.

Recently, a novel approach in the study of accretion powered sources, which
combines variability and spectral information,  has been taken by Revnivtsev,
Gilfanov \& Churazov (1999). Using {\it RXTE} data, they measured  the power
spectrum, and hence the amplitudes of the Fourier components, of Cygnus X-1 in
its hard state for different energies. Then, at each energy band,  they
assembled the Fourier amplitudes  within a given Fourier frequency range, say
$\Delta f$, to produce the so-called Fourier-Resolved (FR) spectrum of the band
$\Delta f$. The process was repeated to obtain the FR spectra of many such
frequency bands, thereby combining the information provided by the time
variability with the simplicity of the insights provided by the ``energy
spectra". Their main conclusions were that: (a) the soft component of the
spectra (thought to represent thermal emission from the innermost parts of the
accretion disk) is absent from the Fourier resolved spectra, indicating that it
is not variable on time scales less than $\sim 100$ s, (b)  The FR spectra
(in the frequencies that are determined) are power-laws which become
progressively harder with increasing Fourier frequency, and (c) the Fe
K$\alpha$ line and reflection components become less pronounced as the Fourier
frequency increases.

Using the same approach, the X--ray continuum spectral variability of the
Galactic black hole binaries (GBHs) GX 339-4 and 4U 1543-47, during its 2002
outburst, as well as  Cyg X-1 in its soft state  have also been studied
respectively by Revnivtsev, Gilfanov \& Churazov (2001); Reig \etal (2006), and
Revnivtsev, Gilfanov \& Churazov (2000). The Fourier-resolved spectroscopy
(FRS) has also been used to study the nature of the  quasi-periodic
oscillations in neutron stars (Gilfanov, Revnivtsev,  \& Molkov 2003, Gilfanov
\& Revnivtsev 2005; Sobolewska \& $\dot{\rm Z}$ycki 2006).

Recently, Papadakis, Kazanas, \& Akylas (2005) applied for the first time the
same technique to an AGN, namely to the \xmm\ observations of MCG -6-30-15.
Their results were similar to those of Revnivtsev \etal\ (1999) in the case of
Cyg X-1 in its hard state,  with the exception of the soft excess component at
energies $E \ltsimeq 1$ keV which was present in the spectra of all Fourier
bands, implying variability of this component in all frequency bands examined,
in  contrast with the behavior of Cyg X-1.

In the present note we apply the method of Fourier-resolved spectroscopy to  
five more AGN, using observations by \xmm. Our aim is to   explore the spectral
variability properties of a sizable sample  of AGN, as implied by the
application  of FRS, in order to investigate  whether potential common trends
and similarities with GBHs.  In this work we report our results regarding  the
AGN spectral variability properties at energies above 3 keV. In this energy
band, the AGN time-average spectra are dominated by a power-law continuum.
Reflection features like the iron K$\alpha$ line, and the associated absorption
edge, appear as well. At lower energies, many AGN show considerable complexity,
caused either by the presence of warm absorbing material and/or by the presence
of the so-called soft-excess emission component. In principle, the warm
absorber can respond to variations of the underlying continuum while the soft
excess emission is often observed to be  variable. Consequently, FRS can be
used in order to study in detail their variability properties. We plan to
present the results from such a study in the near future.

In \S 2 and \S 3 we discuss  the data sets and the methodology we use,
respectively. In  \S 4 we present the results from various model fits to the
time average and  FR spectra of the objects in our sample. Finally, in \S 5 and
\S 6 we discuss our results briefly in the context of theoretical ideas and
models presently in the literature and we present our summary and future plans,
respectively.

\section{Observations \& Data Reduction and Analysis Method}

\xmm, with the large effective area of its instruments  and its capability to
observe a source  continuously for a period up to $\sim 1.3$ days, can provide
long duration light curves that are appropriate in order to obtain  accurate
Fourier-resolved spectra for the AGN. To this end, we searched the \xmm\ public
data archive for those AGN which have been observed, at least once, for an
on-source exposure time larger than 100 ks. Prior to April 2005, there were 5
AGN which satisfied this criterion, namely \mrk,  NGC~3516, NGC~3783, NGC~4051,
and \ark. For those objects, we also considered all the available observations
in the archive, as in principle we can combine the data from different
observations in order to estimate the FR spectra as accurately as possible.

Apart from the criterion regarding the exposure time, we also required  that
the objects show significant variations in all the energy bands that we
consider (see below). For this reason, we used 1000 s binned light curves  (in
this way the source counts in each bin were, on average,  larger than 20) and
applied the usual $\chi^{2}$ test. We found that  all  sources displayed
variations significant at more than the 95\% confidence level, in all energy
bands, except  from the November 2001 observation of NGC~3516. Significant
variations can be detected during the  April 2001 observation of this source.
Although the on-source exposure time in this case  is smaller than 100 ks, we
decided to keep the source in our sample and  use the data from this
observation to study its FR spectra.  As for the  shorter observations of the
other sources, none of them showed significant  variations, except from the
November 2002 observation of  NGC~4051. In fact, the time-average spectra of
the source during the 2001 and 2002 observations are so different that  we
decided to study the two FR spectra separately.

In Table~1 we list the details of all the observations that we have used in
the present work. All data have been reduced using {\footnotesize XMMSAS
v6.1.0}. We use data from the European Photon Imaging Camera ({\footnotesize
EPIC}) pn  detector only. All sources were observed on-axis. With an average
count rate of less than $\sim$ 30 cts~s$^{-1}$ in all cases, photon pile-up
is   negligible for the PN detector, as was verified using the {\footnotesize
XMMSAS} task $epatplot$. Source counts were accumulated  using a circular
region  of $40''$  around the  position of the source. Background data were
extracted from a similar size, source free, region on the chip. We selected
single and double pixel events ({\footnotesize PATTERN$=0-4$}) in the energy
range from 200 eV to 10 keV.

The background was in general low and stable throughout all the observations,
with the exception of short periods at  the start and/or at the end of each
observation. Data from these  periods  of  high  background levels were
removed. The ``exposure times" listed in the third column of Table~1 refer to
the total on-source exposure time of the pn detector, after these high
background level periods are removed.

In Figures~\ref{fig:lc1} and \ref{fig:lc2} we show the $0.2-10$ keV,
background-subtracted, 100 s binned  light curves, extracted from the data of
all observations that we consider in this work. All light curves are normalized
to the lowest count rate observed. In this way, one can easily judge their
``quality". For example, NGC~3516  shows the smallest amplitude variations
(min-to-max amplitude ratio of $\sim 1.5-1.6)$. Its light curve  is also short,
and of low signal-to-noise ratio. On the other hand,  NGC~4051 shows the
largest amplitude variations (with a  min-to-max ratio of $\sim 15$ during the
2001 observation). The longest light curve is that of NGC~3783. This source is
bright, and also displays  significant variations on all sampled time scales,
with a maximum amplitude of $\sim 2.5$.

\section{Analysis Method}

In this section we present in some detail the theory on which FRS is based, as
this is still a novel analysis method, rarely used in the variability studies
of compact objects. The method is based on the fact that any stationary process
can be represented as the ``sum" of sine and cosine functions (e.g. Priestley
1989). More specifically, let us denote with $X(t)$ a stationary process, i.e.
the time variable emitted flux from an AGN, for example. Then, $X(t)$ can be
represented as,

\[ X(t)=\int_{0}^{+\infty}{\rm cos}(\omega t)  dU(\omega) +
\int_{0}^{+\infty}{\rm sin}(\omega t)  dV(\omega). \]

The integrals above are stochastic and defined in the mean-square sense
(Priestley 1989). As for the stochastic processes $dU(\omega)$ and
$dV(\omega)$, they are orthogonal (i.e. their increments at different 
frequencies are uncorrelated) and, more importantly, for each frequency,
$\omega$, one can write,

\[ \langle|dU(\omega)|^{2}\rangle = \langle|dV(\omega)|^{2}\rangle = h(\omega)d\omega , \]

\noindent where the brackets denote  the mean of a random variable, and
$h(\omega)$ is the (non-normalized) power spectral density function of the
stationary process $X(t)$. In other words, the amplitude of the sine and cosine
(random) functions that can be used to represent $X(t)$ are related to the
power spectral density function of $X(t)$. This is the `crucial' property that
we can use in practice to estimate the amplitude of the sine and cosine
functions (\ie\ the ``Fourier components" of the random process under study).

Suppose we observe the X--ray emission from an AGN $N$ times over a period of
$T$ s. Let us denote with $\Delta t$ the interval of each observation (so that
$N\Delta t=T$), and  with $x(E,t_{i}) (i=1,2,\ldots,N)$ the $N$ points of the
light curve (in units of counts s$^{-1}$). Note that we have assumed we observe
the object at a particular energy band of (median) energy $E$. Using the
discrete Fourier-transform of the light curve, we can estimate the power
spectral density function (PSD) of the random process (whose one realization is
the light curve at hand) as follows,

\begin{equation}
P(E,f_{j}) = \frac{2\Delta t}{N} \left| \sum_{i=1}^{N} x(E,t_{i})e^{-2\pi f_{j} 
t}\right|^{2}.
\end{equation}

The units are (counts s$^{-1}$)$^{2}$ Hz$^{-1}$, and the PSD is estimated at
the set of frequencies, $f_j=j/T, j=1,...,N/2$. Based on what we mentioned
above, the quantity,

\begin{equation}
 R(E,f_{j}) = \sqrt{P(E,f_{j}) \Delta f} \hspace{0.2 cm} {\rm
 (counts\hspace{0.1cm} s^{-1}}),
\end{equation}

\noindent (where $\Delta f=1/T$) can be considered as an estimate of the amplitude of the
Fourier components with frequency $f_{j}$.

Suppose now that we have obtained $N_{E}$ light curves at different energy
bands with median energy $E_{k}, k=1,2,\ldots,N_{E}$. If, for each one, we
estimate the amplitude  of the Fourier components with frequency $f_j$, i.e.
$R(E_k,f_j)$, then the plot of $R(E_k,f_j)$  as a function  of energy,
constitutes the ``energy spectrum of the amplitudes of the Fourier-component
with frequency $f_j$",  or the ``Fourier-resolved spectrum at frequency $f_j$".

Although some of the objects we study in this work are quite bright (for AGN),
it is not possible to study their FR spectra using the full energy resolution
offered by \xmm. Instead, for each object  we extracted light curves in seven
bands from 3 to 7 keV with $\Delta E=0.5$ keV, and also the $7-8$ and $8-10$
keV bands, using a bin size of 100 s. We corrected for the background
contribution, estimated the power spectrum (using equation 1), and then
subtracted the contribution of the Poisson noise.

In all objects, many of the $P(E_k,f_{j})$ values, after the subtraction of 
the Poisson noise, are negative, especially at the highest energies and
frequencies. As a result, the estimation of the respective $R(E_k,f_{j})$ is
not possible. For this reason,  we considered two  frequency ranges, namely
$10^{-5}-5{\times}10^{-4}$~Hz  and $5{\times}10^{-4}-1{\times}10^{-3}$~Hz
(hereafter the ``LF" and ``HF" bands). First we  estimated the average of the
power spectrum estimates in each band, and then we used equation (2) to
estimate the average amplitude of the Fourier components in the respective
band.

The frequency ranges that we consider are very broad. The LF and HF bands
correspond to Fourier components with periods from $10^{5}$ to  $2\times
10^{3}$ s  and $2\times 10^{3}-1000$ s,  respectively. This is necessary in
order to estimate, as accurately as possible, the average amplitude of the
Fourier components, especially in the higher energy bands. However, this choice
unavoidably affects the accuracy of the errors of the FR spectra. The error
estimation of the average Fourier amplitudes is based on the  scatter of the
individual amplitudes around their mean in each frequency band. However,  due
to the red noise character of the AGN power spectra, this scatter is
representative, to some degree, of the intrinsic, power-law like dependence of
the Fourier amplitudes on frequency. Consequently, the estimated errors of the
FR spectra are expected to be overestimated. This effect should depend on the
power spectrum slope of each individual source, and is expected to affect more
the LF band estimates. In \S 4 we discuss how we addressed this issue during
the FRS model fitting process.

\subsection{Interpretation of FRS}

The past few years, plots of the energy-dependent variance, $\sigma^{2}_{E}$,
or the rms fractional variability, \ie\ $r_{E}=\sigma_{E}/\langle x_{E}
\rangle$ (where $\langle x_{E} \rangle $ is the average count rate)  as a
function of $E$ have become increasingly popular in the study of the  spectral
variability properties of AGN and GBHs (see for example Edelson \etal\ 2002,
Taylor, Uttley \& McHardy 2003, and Markowitz, Edelson \& Vaughan 2003 for the
application of this method in X--ray variability studies of AGN).  Roughly
speaking, the variance is equal to the integral of the power spectrum of the
source, i.e. $\sigma^{2}_{E}=\int_{1/T}^{\infty}P(E,f) df$, where $T$ is the
length of the observed light curve.  In practice, this integral can be
approximated  by the sum: $\sum_j  P(E,f_{j}) \Delta f = \sum_j
R^{2}(E,f_{j})$.  Consequently, the  ``rms vs. $E$" plots and FRS are related
analysis methods.

In effect, FRS ``decomposes" the rms in each energy band into the contribution
of the individual Fourier components.  Obviously, the FR spectra provide 
``more" information, in the same way that a power spectrum provides ``more"
information than just the variance of a light curve. However, while the
variance can be estimated ``easily",  the requirements for an accurate estimate
of the power spectrum are much more demanding. The same is true for FRS and
the``rms vs $E$" plots. We need longer, and high signal-to-noise light curves
in order to perform Fourier-resolved spectroscopy, while a rough estimate of
the ``rms vs $E$" plot can be achieved with lower quality data.

Although the units of the Fourier-resolved spectra are the same as those of the
observed energy spectrum, they {\it cannot} be interpreted in the same way. 
While the energy spectrum exhibits the distribution of the emitted flux as a
function of energy, the Fourier-resolved spectrum provides the {\sl amplitude
of variability} in a certain frequency range, say $\Delta f$, as a function of
energy. Furthermore, while the integral of the energy spectrum over a certain
energy range, say $\Delta E$, is equal to the power emitted from the source
over that energy band, the integral of the Fourier-resolved spectrum is equal
to the contribution of the Fourier components, in the frequency range $\Delta
f$, to the variance of the light curve in the energy band $\Delta E$. 
Consequently, the use of the word ``spectrum" for a ``$R(E,f)$ vs $E$" plot can
be misleading. We will keep using this term though,  as this is what has been 
used in the past and a change of terminology may cause confusion. However we 
emphasize again that the Fourier-resolved spectra do {\it not} show how {\it
photons} are distributed as a function of energy. They simply show how the {\it
variability amplitudes}, at a certain frequency, change with energy.

So, what is the use of these ``spectra" in practice? Their important property
is that  they receive contribution only from the spectral components which are
variable on the time scales sampled by the observations. For example, let us
consider the case of an AGN with a power-law (PL) X--ray continuum of slope
$\Gamma$. Suppose now that apart from this PL component,  other  spectral
components (like e.g. reflection from a cold or ionized disk and/or heavy
absorption by warm material) also appear in the time-average spectrum of the
source. Because of the presence of such components, sometime it is difficult to
determine $\Gamma$. However, if only the PL component varies in normalization,
then, as we show in the Appendix,  the Fourier-resolved spectra will have a
power-law shape  of slope  $\Gamma$, at all frequencies. Hence, in this case,
the FR sectra  can not only show the variable component, but provide also an
accurate estimation of $\Gamma$ as well.

A straightforward utility of FRS is found in the study of  spectral features
that result from reprocessing of the continuum since in this case there exist a
natural filter (the light crossing time) which  filters out all frequencies
higher than $\sim R/c$ ($R$ is the size of the reprocessing area and $c$ the
speed of light). Such a feature could be  an emission line at energy, say
$E_{0}$, produced by continuum reprocessing over a region of size $R$. Should
its normalization vary in proportion to  the underlying continuum at a given
frequency, its EW should remain constant for frequencies $\nu \ltsimeq R/c$
while it would be vanishing for $\nu  \gtsimeq R/c$. Lower EW values at a
certain frequency range will imply that the  line is not ``as variable" as the
continuum on the respective time scales, either because of the light crossing
argument or because the physical condition at the corresponding radius do not
favor the presence of the associated  transition.

In summary, the Fourier-resolved spectra can show us clearly if and, most
importantly, how the various spectral components in the overall energy spectrum
of a source vary on the frequency ranges considered. The easiest way to
accomplish this, in our case, is to perform a standard model fitting analysis
to the FR spectra in the LF and HF bands and then compare the results with
those obtained from a similar analysis of the time-average energy spectrum.
The model fitting to the time-average spectrum can identify the spectral
components which contribute to the emitted radiation from the source. The
results from the model fitting  of the FR spectra will identify {\it which} one
of the individual spectral components is variable. Any differences between the
best fitting parameter values of the time-average spectrum and the LF/HF FR
spectra will give us information as to {\it how} the respective spectral
components vary.

We discuss below the results from the application of this method to the data of
the five AGN we study in this work.

\section{Spectral Analysis \& Model Fits}

The spectral model fits have been performed with the {\footnotesize XSPEC
v11.3} package. The errors on the best-fitting model  parameters that we report
represent the $1\sigma$ confidence limit for one interesting parameter. The
energy of the emission and absorption features are given in the rest frame  of
the source. Since the number of points in the mean energy spectrum and the FR
spectra  is small, whenever possible, we performed the model fitting with the
parameters of the  emission or absorption features kept fixed at ``sensible"
values (\ie\ at 6.4 keV for the iron K$\alpha$ line and 7.1 keV for the
associated absorption edge).

We consider a model as providing an acceptable fit to the data if the null
hypothesis probability is larger than 5\%. We accept that the addition of a
model component is necessary if the quality of the model fitting is improved at
more than the 95\% significance level. All spectral fits include Galactic
absorption, with column values taken from Dickey \& Lockman (1990). They are
listed on the top of the Tables where we report our best fitting model
parameter values. 

Spectral responses and the effective area for the pn spectra were generated
with the {\footnotesize SAS} commands {\em rmfgen} and {\em arfgen}. Since both
the time-average and the FR spectra have a much coarser energy resolution than
the intrinsic resolution of the EPIC pn detector, we  used the {\footnotesize
FTOOLS} command {\em rbnrmf} to rebin the original pn response matrix
accordingly.  Furthermore, a uniform systematic error of 1\% was added
quadratically to the statistical error of the time-average spectra to account
for all the systematic uncertainties that may be introduced when we
undersample  the  original energy resolution of the instrument and use a
``binned" response matrix.

This systematic error was not added to the Fourier-resolved spectra, since as
we mentioned above their errors are probably overestimated anyway. In order to
resolve this issue we followed a model-dependent procedure. We fitted each FR
spectrum with a simple power-law model in the energy bands $3-5.5$ and $7-10$
keV. In most cases, the resulting reduced \chis\ values, \chir,  were
significantly smaller than 1. We would then reduce  the errors by an
appropriate factor (equal to $\sim 2-5$ and $\sim 1.5-4.5$ in the case of the
LF and HF spectra, respectively) so that \chir=1. The  resulting
error-correction factors were then applied to the $5-7$ keV band points as
well.

\subsection{Mrk~766}

The time-average spectrum and the LF/HF FR spectra of \mrk\ are plotted in the
upper panel of Figure~\ref{fig:mrk} (open circles, filled squares and filled
triangles, respectively). The mean spectrum is well fitted by a power law model
of $\Gamma_{\rm av}\sim 2.15$ and a broad Gaussian line with   $E_{\rm
line,av}\sim 6.45$, $\sigma\sim 480$ eV and EW$_{\rm av} \sim 230$ eV. The best
fitting model is shown with the dashed line in the upper panel of
Figure~\ref{fig:mrk} and the best fitting model parameter values are listed in
Table~2. They are in good agreement with the results from the same model fit to
the full energy resolution EPIC pn spectrum (see \S 3.2 in Pounds \etal\ 2003).

In the lower three panels of the same Figure we plot the Data/Model ratio  in
terms of ``sigmas" (\ie, the error of each point; as a result, the errors of the
points in these plots are of size one). In the case of the time-average spectrum,
``Model" refers to the best fitting model (with the parameter values listed in
Table~2) while in the case of the LF and HF spectra, as ``Model"  we use the best
PL model fit.

The HF spectrum appears to be rather noisy but a  PL model ($\Gamma_{\rm
HF}\sim 1.9$) fits it rather well. The LF spectrum is also well fitted by a
simple PL model ($\Gamma_{\rm LF}\sim 2.2$, $\chi^{2}/$degrees of freedom
(dof)=11.4/8). However, when we add a narrow Gaussian line (\ie\  $\sigma$ kept
fixed at 100 eV) with $E_{\rm line}$ ``frozen"  at 6.45 keV (the best fitting
value in the case of the time-average energy spectrum) we find that
$\chi^{2}/$dof =5.5/7. According to the F-test, the addition of the narrow line
is significant at the 97.1\% level.

\subsection{NGC~3516}

The time-average energy spectrum and the LF FR spectra of NGC~3516 are plotted
in the upper panel of Figure~\ref{fig:3516} (open circles and filled squares,
respectively). The $3-10$ keV mean energy spectrum is very flat. This can be
explained by  the presence of high column layers of absorbing material in
various ionization states (Turner \etal\ 2005). A simple PL model cannot
provide an acceptable fit to it. For that reason, we restricted our model
fitting to the $4-10$ keV band. We found that a flat PL ($\Gamma_{\rm av}\sim
1.35$; see Table~3) plus a narrow Gaussian line with the centroid energy kept
fixed at 6.4 keV (shown with the dashed line on the top panel in
Figure~\ref{fig:3516}) can fit the time-average spectrum well.

Since this source shows the smallest amplitude variations, and has the lowest
signal-to-noise ratio among the sources in our sample, it is not possible to
estimate its HF spectrum. At high frequencies, the intrinsic variations in
almost all energy bins are lost in the strong Poisson noise signal.  It was
possible though to estimate the LF Fourier-spectrum. It is well fitted with a
power law model with $\Gamma_{\rm LF}\sim 1.34$. There does appear a small
amplitude, positive excess at $E\sim 6.5-7$ keV in the residuals plot, however
the addition of a narrow Gaussian line component to the PL model does not improve
significantly the goodness of the model fit in this case.

\subsection{NGC~3783}

NGC~3783  was observed  by {\it  XMM-Newton} for  two  complete orbits between
2001 December 17 and  2001 December 21, producing a total good data exposure
of  248 ks. Gaps in the observations  due to a telemetry drop-out during the
first  orbit, and targeting restrictions between the first  and  second
orbits,  forced  us to  analyze  the  data as  three separate observations.
Once the Fourier amplitudes  of each separate observation segment were
calculated, we proceed by combining them in  order to calculate the
Fourier-resolved spectra for the  whole observation.

The energy spectrum of the source above 3 keV is quite complex (Reeves \etal\
2004). The time-average spectrum (shown with open circles in the upper panel of
Figure~\ref{fig:3783}) is well described by a PL component, together with an
iron line at 6.4 keV and an absorption edge at 7.1 keV. Our estimated EW$_{\rm
av}$ of the line ($\sim 100$ eV) is in good agreement with the  Reeves \etal\
(2004) estimate of $107\pm 8$ eV (as listed in their Table 1). Our best fitting
power law slope though ($\Gamma_{\rm av}\sim 1.5)$  is harder than their
estimate of $\sim 1.7$. This is probably caused by the fact that they have
considered a combination of a PL plus a cold reflection model ({\it pexrav} in
{\footnotesize XSPEC}) which naturally results in a steeper PL slope. Due to
the reduced energy resolution ( i.e. the small number of points in the FR
spectra), we cannot really add a reflection component in the modeling of the
time-average energy spectrum, as in this case, the best fitting model
parameters are essentially unconstrained.

Just like in the case of NGC~3516, despite the availability of long, high
signal-to-noise light curves, it was not possible to estimate accurately the HF
spectrum of the source. We were able to estimate the LF spectrum though.  The
lower panel in Figure~\ref{fig:3783} shows the residuals from the best fitting
PL model to the LF spectrum. A PL model does not provide an acceptable  fit to
the $3-10$  keV LF spectrum ($\chi^{2}=$ 29.9/8dof). Clearly, a line-like and
edge-like feature at  $\sim 6$ keV and $\sim 7$ keV, respectively, are present
in the residuals plot. These features strongly suggest the presence of a
variable reflection component, at least on time scales $ 10^{5}-2\times 10^{3}$
sec. Indeed the addition of a narrow Gaussian results in a statistically
acceptable fit (the best fitting results are listed in Table~2).

\subsection{NGC~4051}

We have analyzed two archived XMM observations of NGC~4051. One was taken in
2001, May 16 and the other in 2002, November 22. The source was in a
particularly low flux state during the second observation (Uttley \etal\ 2004;
Pounds \etal\ 2004). The 2001 and 2002 time-average and Fourier-resolved
spectra are shown in Figures~\ref{fig:4051-1} and \ref{fig:4051-2},
respectively. The best fitting results for the 2001 and 2002 spectra  are
listed in columns 2-4 and 5-6  of Table~5, respectively.

The time-average spectrum during the 2001 observation is well fitted by a
$\Gamma_{\rm av}\sim 1.8$ PL model plus a narrow Gaussian line at $\sim 6.1$ keV
(EW$_{\rm av} \sim 85$ eV) and a 7.1 keV absorption edge ($\tau_{\rm av} \sim
0.1$). Our results are in agreement with those presented in \S 3.1 and 3.2 of
Pounds \etal\ (2004).

A simple PL model yields a rather poor fit to the 2001 LF spectrum (\chis
=16.2/8 dof, null hypothesis probability 4\%).  The respective residuals plot
in Figure~\ref{fig:4051-1} reveal a positive excess at $6-7$ keV and a deficit
at $\sim 7-8$ keV. These features are suggestive for the presence of a
reflection component. When we add a narrow Gaussian line, with E$_{\rm line}$
fixed at the time-average spectrum best fitting value, the fit is now
``acceptable", \ie, the null hypothesis probability is now $>5$\%. On the other
hand, the HF spectrum is well fitted by a simple PL model, with $\Gamma_{\rm
HF}\sim 2$.

The 2002 time-average spectrum is quite complicated. It is very flat
($\Gamma_{\rm av}\sim 0.9$), a fact that Pounds \etal\ (2004) explained in
terms of partial covering of the central source by ionized material while
Uttley \etal\ (2004) suggested that the reflection component in 2002 is
stronger than that during the 2001 observation. A power-law plus a narrow
Gaussian line (with a flux of $\sim  1.5\pm 0.02 \times 10^{-5}$ photon
s$^{-1}$ cm$^{-2}$, in agreement with the Pounds \etal\ measurement) and a 7.1
keV edge does provide a good fit to the 2002 time-average spectrum.

Due to the short duration of the 2002 observation, the HF Fourier-resolved
spectrum could not be determined accurately. The LF spectrum on the other hand
is well fitted by a simple PL model. Interestingly, the best-fitting PL slope
($\sim 1.8$) is much steeper than that of the time-average spectrum.

\subsection{\ark}

\ark\ is an X--ray bright, Narrow Line Seyfert I galaxy, which exhibits large
amplitude variations on short time scales. Its time-average spectrum (plotted
with open circles in the upper panel of Figure~\ref{fig:ark}) is well fitted by
a steep power law ($\Gamma_{\rm av}\sim 2.5$) plus a narrow, weak (EW$_{\rm
av}\sim 85$ eV) Gaussian line at $\sim 6.65$ keV (see Table~6). These results
are in good agreement with those of Papadakis \etal\ (2007), who have studied
the full-resolution pn spectrum of the source.

Its Fourier-resolved spectra are rather simple in shape. They are well fitted
by  PL models (the best fitting results are listed in Table~6). The FR spectra
best fitting slope values are comparable to the best fitting PL slope of the
time-average spectrum. The LF residuals' plot is somehow noisy, however, the
addition of a narrow Gaussian line at $\sim 6.7$ keV does not improve
significantly the goodness of fit of the PL model.

\section{Discussion}

We have applied the Fourier frequency-resolved spectral analysis method to
\xmm\ data of five AGN  with the main aim of studying their spectral
variability. This work supplements earlier analysis of  the \xmm\ data of MCG
6-30-15 by Papadakis \etal\ (2005) and work on Galactic sources by a number of
other authors (see Introduction). We find that  AGN present a much larger
variety in the properties of their FR Spectra compared to the (still limited in
number) FR spectra of GBHs. We see several obvious reasons for that. First of
all, even our longest observation ($\sim 200$ ks) covers a much smaller range,
in terms of the objects' characteristic frequency, than the several hours of
data of a Galactic source (see discussion in later on in this section).
Furthermore, there is evidence that the AGN emission is reprocessed over a
larger range of radii than that  of Galactic sources, a feature that
complicates considerably the comparison between  time averaged and FR spectra
in AGN. Nevertheless, when we consider the results from the FRS analysis of all
the sources together, we can identify some common trends which can be
summarized as follows:

1) Both the LF and HF Fourier-resolved spectra of all sources are well fitted
by power-law models. We also find that, in  general, there are differences
between the spectral slopes of  the time-average, the LF and the HF spectra.
The  interpretation of this fact is not apparent at this point; some
alternatives are given later on.

2) A line-like feature at energies $\sim 6-6.5$ keV appears the LF spectra of
\mrk, NGC~4051 and NGC~3873. The most straightforward explanation of these
facts is that there is a component of the iron line in  these objects that is
variable, to some degree, on time scales of the order of a few hours up to a 
day. 

3) We do not detect Fe line features in any of the three HF spectra  that we
could estimate. This implies the absence of variability of  this feature on
time scales less than a few hours, despite the  significant continuum
variations on the same time scales. 

\subsection{The FRS slope vs frequency relation}

The fact that the spectral slopes of the time-average, LF and HF spectra are
not always the same implies that the continuum PL in these objects does not 
vary just in normalization. However, if we consider them individually, it is
not easy to interpret the results regarding the slope of the Fourier-resolved
spectra ($\Gamma_{\rm FRS}$). 

For example, in \mrk\ and in the 2001 observation of NGC~4051 we observe a
$\Delta \Gamma\sim 0.2-0.25$ difference between the slope of the LF and HF
spectra  ($\Gamma_{\rm LF}$ and $\Gamma_{\rm HF}$, respectively, with the
latter being harder). In both cases, the (absolute) difference between the FRS
and the time-average spectral slope ($\Gamma_{\rm av}$) is rather small, and of
the order of $|\Delta\Gamma| \sim 0.1-0.3$. 

The difference $\Gamma_{\rm LF} -  \Gamma_{\rm av}$ in the case of NGC~3783 and
of the 2002 observation of NGC~4051 is even larger  ($\Delta\Gamma \sim 0.5$
and $\sim 1$, respectively). In both cases, the slope of the LF spectrum
($\Gamma_{\rm LF}\sim 1.8-2$)  does make sense, as it is close to the
``canonical" slope of the AGN continuum spectrum, suggesting  that the
intrinsic slope of their X--ray continuum spectrum is indeed $\Gamma \sim
1.8-2$. This result shows clearly the benefits of the  Fourier-resolved
spectroscopy. The fact then that their time-average spectrum appears to be much
flatter must be  caused by external factors like absorption by ionized material
and/or the substantial contribution from a reflection component. 

On the other hand the LF spectrum of NGC~3516 has a slope similar to that of
the time-average spectrum of the source (which is harder than the ``canonical"
value of $\Gamma \sim 1.9-2$). The same result holds for the LF and HF spectra
of \ark. They are both as steep as the time-average spectrum of the source.

The picture becomes clearer if we consider all the results together  and plot
the FR spectral slope as a function of frequency. Since the central source of
each AGN in our sample has a different Black Hole (BH) mass, the LF and HF
bands correspond to different intrinsic time scales in each system. To this
end, we assumed that the LF and HF best PL fitting slope values are
representative of the Fourier-resolved spectral slope at the mean frequency of
the LF and HF bands ($2.55\times 10^{-4}$ Hz and $7.5\times 10^{-4}$ Hz,
respectively). Then, for each object, we divided this frequency with its
Keplerian frequency at 3R$_{S}$, $f_{\rm K}(3\rm R_{S})$, where  R$_{\rm S}$ is
the Schwarzschild radius. BH mass estimates for NGC~3783, NGC~3516 and NGC~4051
were taken from Peterson \etal\ (2004). For \ark\ and \mrk, we used the
$2.6\times 10^{6}$ and $3.5\times 10^{6}$ M$_{\odot}$ estimates of Botte \etal\
(2004) and Woo \& Urry (2002), respectively.

Figure~\ref{fig:gamma} shows a plot of  $\Gamma_{\rm FRS}$ as a function of
$f_{\rm norm}$, \ie\  the frequency normalized as explained above (filled
squares, except for \ark\ and  NGC 3783 which are shown separately in the
Figure). In the same set of points we have also included the results from the
Papadakis \etal\ (2005) study of MCG-6-30-15. In this case, the mean of the
three frequency bands that these authors had considered were normalized to the
Keplerian frequency at a distance of 3R$_{\rm S}$ from a BH of mass $4.5\times
10^{6}$ M$_{\odot}$ (McHardy \etal\ 2005). Clearly, taken as a whole, the AGN
results suggest that the slopes of the Fourier-resolved spectra harden with
increasing frequency. A model of the form  $\Gamma_{\rm FRS}\propto {\rm ln}
f_{\rm norm}^{\alpha}$ describes rather well the AGN data, with $\alpha\sim
-0.25$ (solid line in Figure~\ref{fig:gamma}).

In the same Figure, we also plot the results of Revnivtsev \etal\ (1999) and
Gilfanov \etal\ (2000) for Cyg X-1 in its hard/low (LS) and soft/high (HS)
state  (open squares and open circles, connected with a dashed line,
respectively). We have assumed a BH mass of 10 M$_{\odot}$.  The long-dashed
line in Figure~\ref{fig:gamma} shows the $\Gamma_{\rm FRS}$ vs $f_{\rm norm}$
best fitting model for the AGN shifted to lower $\Gamma$ values, appropriate
for an extrapolation of the Cyg X-1 data in LS to higher values of  $f_{\rm
norm}$. At this point, one could consider that, at high frequencies,  the
$\Gamma_{\rm FRS}$ vs $f_{\rm norm}$ relation is similar both for the AGN we
study and Cyg X-1 in its LS,  but shifted by $\Delta \Gamma \sim 0.5$.

The \ark\ results (plotted with open diamonds in the same Figure) appear to be
consistent with those of Cyg X-1 in its HS. In this state, the  $\Gamma_{\rm
FRS}$ remains constant up to frequency $\sim  0.03 f_{\rm K}(3\rm R_{S})$ (\ie,
ten times higher than Cyg X-1 in its LS), and then the Fourier-resolved spectra
flatten. Consequently, it is our view that  the \ark\ $\Gamma_{\rm FRS}$ are
similar to the time-average spectral slope simply because the  system operates
at a spectral state different than that of the other AGN in our sample and
similar to the HS of Cyg X-1.  In this latter case, one would have to probe the
FR spectra at frequencies higher  than those probed to-date in order to observe
the decrease of $\Gamma_{\rm FRS}$  with increasing frequency.  The NGC~3783 LF
slope is also consistent more with  the Cyg X-1 data in HS. In fact, the
strength of the Fe line in its LF spectrum suggests that this object may
indeed  be operating like Cyg X-1 in HS (see below).

The dotted line in Figure~\ref{fig:gamma} shows the  $\Gamma_{\rm FRS}$ vs
$f_{\rm norm}$ best fitting model for the AGN, shifted to higher $\Gamma$
values, appropriate for the \ark\ data. One can see that the NGC~3783
$\Gamma_{LF}$ value consistent with this line, which also seems to be
appropriate for an extrapolation of the Cyg X-1 data in HS to higher values of
$f_{\rm norm}$.

When we combine together the results from  Cyg X-1 and  the AGN that we present
in this work, the following picture for the accreting black hole systems
emerges: a) $\Gamma_{\rm FRS}\sim \Gamma_{av}$, at all frequencies, up to a
certain value, say $f_{\rm br}$, b) $f_{\rm br}\sim 0.003\times f_{\rm K}(3\rm
R_{S})$ and  $f_{\rm br}\sim 0.03\times f_{\rm K}(3\rm R_{S})$ for systems in
their  Low/Hard and High/Soft state, respectively, and c) at frequencies $f >
f_{\rm br}$, $\Gamma_{\rm FRS}$ decreases with increasing frequency as ${\rm ln} 
f_{\rm norm}^{-0.25}$, for all systems, irrespective of the state they operate.

We believe that the $\Gamma_{\rm FRS}$-vs.-$f_{\rm norm}$ relation that we
present in this work provides new, important clues as to  the origin of
X--ray spectral variability in AGN. From a phenomenological point of view,  our
results suggest that the power-law like continuum in AGN does not vary in
normalization only. If that were the case, we would expect $\Gamma_{\rm FRS}$
to be constant with $f_{\rm norm}$. We conclude that intrinsic spectral slope
variations must occur, although their pattern is not clear at the moment. For
example, as we show in the Appendix,  $\Gamma_{\rm FRS}$ should not change with
$f_{\rm norm}$ even in the case of a pivoting power-law (when the pivot energy
is outside the energy range sampled by the observations). Therefore, simple
phenomenological models like that of a pivoting energy spectrum can not 
provide an obvious interpreation of our results.

In general, most models that are available at present do not address  the
constrains that the Fourier-resolved spectroscopy imposes on the X--ray
production mechanism in black hole accreting objects. One exception is 
$\dot{\rm Z}$ycki (2003) who calculates the FR spectra (and compares them with
what is observed in Cyg X-1) in the case when  X--ray emission is attributed to
active regions moving radially towards the central compact object. In this
model, the Comptonized X--ray spectrum from each region is assumed to evolve
from softer to harder during the flare evolution as result of diminishing
supply of seed photons due to the fact that either the disk is absent at
smaller radii or the thickness of the disk's ionized skin increases toward the
centre. The model can produce FR spectra whose slope hardens with increasing
frequency in a  way similar to what we observe in GBHs (and AGN, as we show in
this work) assuming the presence of a wide range of infall velocities, and
that each such region stimulates the production of other regions as well, in a
fashion similar to the ``avalanche" model of Poutanen \& Fabian (1999) with
faster variations appearing later in the spectral evolution of these regions,
along with the hardening  of their spectra. 

However, the predictive power of this model is somehow  hampered by the fact
that it involves a large number of parameters whose values cannot be predicted
in a physical way. For example, one has to assume rather arbitrary if and how
the velocity of the regions depends on radius, the maximum/minimum infall
velocities, the average value of regions that exist  per unit time, the
probability that an active region will activate a second one etc. Perhaps the
$\Gamma_{\rm FRS}$-vs.-$f_{\rm norm}$ relation that we present in this work can
constrain meaningfully the model parameter values and hence help us undertand
better their physical implications.

On the other hand, a qualitative account of the FRS results can also be
obtained in the simpler  case of Advection Dominated Accretion Flows (ADAF;
Narayan \& Yi 1994), provided that one is willing to accept a relation between
the Fourier frequency and the size of the emitting region. In ADAF, for a wide
range of accretion rates, the electron temperature attains values typically
$\sim 100$ keV, and  remains roughly constant in radius for $x<10^2-10^3$,
where $x$ is the radius  normalized to $\rm R_{S}$. Since the local density,
$n(x)$, changes with radius as $x^{-3/2}$, the Thompson depth of the flow at
radius $x$, $\tau_T(x)$, should vary as $\tau_T(x) \propto x^{-1/2}$.
Consequently,  the Comptonization parameter of the flow, $y(x) = \tau_T(x)
(kT_e/m_ec^2)$, increases with decreasing radius.

The response of the ADAF to a variation of a particular duration is the
convolution of the ACF of the particular variation with the response  function
of the system. Assuming that the response of the system extends over the entire
radial (and hence time) range of the ADAF, a short variation has support mainly
in the small values of $x$, or equivalently the large values of the
Comptonization parameter $y$ of the flow. Alternatively,  a long duration flare
has support over much a larger region of ADAF where the values of $y$ are
smaller, the resulting variation is weighted more by the softer photons leading
to softer FR spectra. Put it in a different way, short variations  produce
harder spectra than the long ones.

 The above are admittedly qualitative arguments. They are put forward to
provide a flavor of the conclusions one can infer from the results of FRS. As
such, they seem to provide some support for models which involve hot flows
(i.e. ADAF-like geometry) or multiple active regions/perturbations which
propagate towards the central object over, say, that of a uniform corona.

\subsection{The Iron Line EW vs. Frequency Relation}

The lack of response, on short time scales, of the reflection spectrum
(including the iron line) on the observed X--ray continuum luminosity variations
in AGN has been a long-standing issue (Iwasawa et al. 1996, 1999, Lee et al.
2000, Vaughan \& Edelson 2001, Markowitz \etal\ 2003, Vaughan \& Fabian 2004).
The present work demonstrates clearly that Fourier-resolved spectroscopy is
``sensitive" in detecting iron line variability. Our results suggest that, at
least in  three out of the five sources in our sample, the iron line is
variable on time scales as short as a few hours. 

Admittedly, the FRS error-reduction method (see \S 3) may affect the
significance of the line-like features that we detect in the LF spectra of
\mrk, NGC~3783 and NGC~4051. However, in the case of NGC~3783, the PL best
fitting residuals' features around $\sim 5-7$ keV are so strong that, even if
we do not reduce the LF errors, the addition of a narrow Gaussian line improves
the PL fit at almost the 99\% significance level. In \mrk, Miller \etal\
(2006), using a much longer, recent \xmm\ data set have shown that the  line
flux is indeed variable, in response to the continuum variations, on
time-scales of a few ksec. It is only in NGC~4051 that the line detection in
its LF spectrum is not highly significant (the addition of a Gaussian line to
the PL model is not statistically required; it simply provides a model fit with
a null hypothesis probability larger than 5\%). Nevertheless, based on the
consistency of the NGC~4051's results with those from the other sources (and
Cyg X-1) that we present below, we believe that the signal we detect in the LF
spectrum of this source is also real.

As with the slope of the FR spectra, it is hard to understand the results 
regarding the EW of the lines we detect in the three sources if we consider
them individually. For example, the EW of the line in the LF spectrum of
NGC~3783 ($200^{+69}_{-73}$ eV)  is larger than that in the time-average
spectrum of the source ($100\pm 16$ eV). This is strange, as in the simplest
case, when both the reflective material and the observer receive the same
amount of primary radiation, the maximum value of the line's EW in the FR
spectra, in any frequency band, should be no more than  the line's EW in the
time-average spectrum. This would suggest that  the iron line fully responds to
the continuum variations, at all frequencies. One could conceivably reconcile
these observations by postulating that the average continuum that is determined
by the time-average energy spectral studies includes also the  contribution
from other components  not seen by the ``cold" matter responsible for its
reprocessing to Fe-line photons.  In this case, the estimated EW will be
smaller than the intrinsic value, which should be at least as large as $\sim
200$ eV.

In \mrk, the EW of the line  is smaller in the LF than the time-average
spectrum (implying that the line is not ``as variable as the continuum", on
time scales of $\sim 1$ hour$-$1 day) while the situation is less clear in the
case of NGC~4051, because of the large uncertainty in the determination of the
LF line's EW.

Just like with $\Gamma_{\rm FRS}$, in order to get a deeper insight as to  how
the iron line varies in AGN, we plot in Figure~\ref{fig:ew} the EW of the Fe
line of the LF FR spectrum, EW$_{\rm FRS}$, for the three objects with
positive  detection  as a function of the Fourier frequency, normalized as
explained above. In the same Figure we also plot the results of Revnivtsev
\etal\ (1999) and Gilfanov \etal\ (200) for Cyg X-1 in LS and HS (open squares
and open circles, respectively).

When Cyg X-1 is in its LS, EW$_{\rm FRS}$ remains constant up to frequencies
$\sim 0.003$ of $f_{\rm K}(3\rm R_{S})$. At higher frequencies, EW$_{\rm FRS}$
decreases. The dashed line in Figure~\ref{fig:ew} shows that a model of the
form EW$_{\rm FRS} \propto {\rm ln} f_{\rm norm}^{-\beta}$ (with $\beta\sim
0.14$)  describes rather well the EW$_{\rm FRS}$ vs $f_{\rm norm}$ relation in
the case of the Cyg X-1 in LS. Clearly, the NGC~4051 and \mrk\ results are
consistent with the extrapolation of this line to higher frequencies. On the
other hand, the surprisingly large EW$_{\rm FRS}$  of NGC~3783 can be explained
if the system operates in the same mode as Cyg X-1 in its high state. In this
case, EW$_{\rm FRS}$ remains constant at all Fourier-resolved spectra up to
$\sim 0.05$ of $f_{\rm K}(3\rm R_{S})$, above which it starts to decrease.

There are however some issues which complicate the analogy between the Cyg X-1
and AGN results. The most puzzling one is the case of \ark. The $\Gamma_{HF}$
and $\Gamma_{LF}$ values suggest that this source operates in a mode similar to
Cyg X-1 in its HS (see discussion in the previous section). If true, we should
then  expect to detect the iron line in its FR spectrum with a high EW, which
is certainly not the case. Recently, Arevalo \etal\ (2006) suggested that \ark\
operates like GBHs in their ``Very High State". Since the relation between
EW$_{\rm FRS}$ and $f_{\rm norm}$ is not known for systems in this state  (the
results of Reig \etal\ (2005) suggest that EW$_{\rm FRS}$  in VHS systems is
smaller than in HS systems), we cannot make any conclusive  statements at the
moment.

The EW$_{\rm FRS}$ vs. frequency relation is amenable to a more straightforward
interpretation than that of the $\Gamma_{\rm FRS}$ vs. frequency relation,
simply because the line emission is the result of reprocessing the X--ray
continuum by neutral, or better, sufficiently little ionized, matter.  Thus
this relation  can be used to probe the reprocessing geometry surrounding the
X--ray emitting source. The EW$_{\rm FRS}$ dependence on the frequency  should
be determined mainly by the  inverse of the light crossing time between the
X--ray source and the  reprocessing matter. Gilfanov \etal\ (2000), assuming
the simplified geometry of an isotropic point source at a height of $\sim
10$R$_{\rm S}$  above a flat disk, have shown that the decrease of EW$_{\rm
FRS}$ at frequencies higher than $0.003$ and $0.03$ of $f_{\rm K}(3\rm R_{S})$,
for systems in LS or HS, respectively,  can be explained if the distance
between the X--ray continuum and the ``reflective" part of the disk is $\sim
100$ and $\sim 10$ R$_{S}$, respectively. $\dot{\rm Z}$ycki (2003, 2004) has
also considered the variability properties of the iron K$\alpha$ line in
accreting black holes. He has found that the EW of the line decreases  with
increasing frequency, in agreement with the results of the Fourier-resolved
spectroscopy, even in the case when there is not just one, static source  but
many,  X--ray producing flares/active regions propagating inwards,  when the
optically thick, cold disk disappeares, gradually, at a distance larger than
the radius of the  innermost stable orbit, and is replaced by a hot, and less
reflective, hot flow.

Interestingly, the normalized frequencies above which EW$_{\rm FRS}$ decreases
with increasing frequency, are very similar to the frequencies above which
$\Gamma_{\rm FRS}$ hardens with increasing frequency. Perhaps then, a
combination of a geometrically thin, optically thick disk (which can reflect
X--rays) and a hot ADAF, with a transition radius of $\sim 100$ and $\sim
10-20$R$_{\rm S}$, in the LS and HS systems, can explain, qualitatively, both
the $\Gamma_{\rm FRS}$ and   EW$_{\rm FRS}$ vs frequency relations we observe.

The fact that both EW$_{\rm FRS}$ and $\Gamma_{\rm FRS}$ decrease with $\nu$
implies a correlation between EW$_{\rm FRS}$ and $\Gamma_{\rm FRS}$ as well: at
frequencies higher than $0.003-0.03$ of $f_{\rm K}(3\rm R_{S})$, the equivalent
width of the line in the FR spectra decreases together with  their slope. To
the extent that EW can be considered a proxy (for the  unobserved) hard X--ray
reflection  amplitude, $R$, this relation appears to be similar to the  $R 
-\Gamma$ relation  of Zdziarski \etal\ (1999). Despite the similarities though,
one has to take into account that our relation is between ``R($\nu$)" and the
slope of the variability amplitudes (at $\nu$) as a function of energy. Our
analysis therefore involves the additional timing  information not present in
the analysis of Zdziarski \etal\ (1999), who report the relation between
quantities determined in the time-average energy spectra.

It is not clear how the $R(\nu) - ``\Gamma"(\nu)$ relation that we find relates
to the $R - \Gamma$ relation  of Zdziarski \etal\ (1999).  These authors
accounted for this result in terms of increase in $\Gamma$ (i.e. softening of
the energy spectrum) due to the increase of the number of cooling photons as
manifest by the increase of the reflection fraction $R$ (assuming implicitly
that the X-ray emission takes place at a geometry fixed for a given source). It
is hard to see how one could produce the results we have obtained within this
framework.

One possibility was described by $\dot{\rm Z}$ycki (2003) in the context of
inwards moving flares whose spectrum evolves (from soft to hard) as they move.
The point is that as a result of irradiation by hard X-rays a hot ionized skin
forms on the surface of the disk. If the luminosity of the flare increases as
it flows inwards, the thickness of the skin increases, the the effectiveness of
reprocessing/thermalization decreases: the EW of the line decreases, soft
photons diminish, and hence the spectra become harder, at the same time that
the variability time scales shorten.

But there are other possibilities as well. As we argued in \S 5.1, and in
Papadakis \etal\ (2005), the decrease in $\Gamma$ with $\nu$ maybe due to the
variation of the Comptonization parameters with scale  (i.e. increasing  $y$
with decreasing distance from the central source), as it is the case e.g. in an
ADAF. The combination of such a flow   attributes with a disk of finite inner
radius, implying that the line photons do not respond fully to the high
frequency variations of the ``harder"  spectra,  could also  then provide a
qualitative account of the correlation of the  ``R($\nu$) vs $\Gamma_{\rm
FRS}(\nu)$" relation.

Testing the above ideas, however, requires a more detailed treatment  of the
reflection process and the way the reflector responds to the continuum
variations. For example, it is important to explain the shape of the EW$_{\rm
FRS}$ vs frequency relation, but also its ``amplitude" as well. The EW of the 
iron line depends on the ionization state of the reflector (Nayakshin, Kazanas
\& Kallman 2000), the solid angle subtended by the reflector, and the abundance
of iron. The increase of  EW$_{\rm FRS}$ with decreasing frequency necessarily 
implies the presence of neutral matter at the corresponding distances
intercepting the solid angle implied by the measured EW$_{\rm FRS}$. A
quantitative  investigation of the necessary physical/geometrical arrangement
that could explain the observed EW$_{\rm FRS}$ values at each frequency
requires  models far more detailed than warranted by the present investigation;
we hope  to return to such models in the future.

\section{Conclusions}

Using long, high signal-to-noise \xmm\ light curves, we estimated the Fourier
frequency-resolved spectra of five AGN. The main results from our study are:

1) The slope of the FR spectra in AGN decreases with increasing frequency (as
$f^{-0.25}$; see Fig.\ref{fig:gamma}) when we take into account properly the
differences in the black hole mass of the central engine in these objects. This
result shows clearly that the PL continuum in AGN must vary in slope, as well
as in normalization, with time.

2) We detect significant evidence in the LF spectrum of three objects (\mrk,
NGC~3783, and NGC~4051) that the iron line is responding to the continuum
variations, on time scales larger than a few hours.

3) We do not detect any evidence of Fe line features in our HF spectra. This
result indicates that the iron line does not respond to the significant
continuum variations on time scales less than an hour.  

We note that, apart from providing answers to questions like  ``is the Fe line
variable?",  Fourier frequency-resolved spectroscopy can show {\it how} the
line emission responds to the continuum variations at certain time scales.  The
data so far suggest that the relation between $EW_{\rm FRS}$ and $f_{\rm
norm}$  (\ie\ $EW_{\rm FRS}$ decreasing with increasing $f_{\rm norm}$) is
similar in AGN and Cyg X-1.

The interpretation of these results is not straightforward. The accounts we
have presented in  \S 5.1 and 5.2 may or may not be correct. Their importance
lies in indicating the potential of the FRS method in probing the  accretion
flow structure of AGN and GBHs. We are encouraged, for instance, by the way
that the NGC~3783 and the 2002  NGC 4051 observation FRS results can be
consolidated within a reasonable, coherent scheme, whereby the intrinsic 
primary PL slope is $\sim 1.8-2$, hence the flatness of the time-average
spectrum is caused by external effects. 

We believe that both the continuation of a systematic study with this method,
but also the effort to develop theoretical models that can describe,
quantitatively, the data as well, have a great potential for providing a more
complete and consistent picture of these objects. Significant progress in the
determination of the  $\Gamma_{\rm FRS}$ and EW$_{\rm FRS}$ vs $f_{\rm norm}$
relation in AGN, and its comparison with the respective relations in GBHs, can
be made with the use of light curves which will allow us to study variations
which  operate on time scales longer than a few days.  In retrospect, if we
look at Figures~\ref{fig:gamma} and \ref{fig:ew}, it seems  rather surprising
that we could estimate $\Gamma_{\rm FRS}$, or detect  line-like features, in
any for the AGN Fourier-resolved spectra that we studied in this work. The
reason is that when compared to Cyg X-1, after taking into account the
difference in the BH mass of the systems, the frequencies we probe in AGN with
the present \xmm\ light curves  are significantly higher than those that have
been studied in GBHs with  {\it RXTE} data.  To this end, the numerous, high
signal-to-noise,  long monitoring light curves in the {\it RXTE} archive will
be helpful.  Their analysis with the FRS method will allow us to determine
accurately the $\Gamma_{\rm FRS}$  and EW$_{\rm FRS}$ relations in AGN. We plan
to perform such a research work in the near future.


\acknowledgements
IEP and ZI gratefully acknowledge support for this work by the General
Sectreteriat of Research and Technology of Greece. DK acknowledges support by a
{\em Chandra} GO grant.

\appendix
\section{Appendix: Calculation of the FR spectra in some simple cases}

According to equation (2) in \S 3, the Fourier resolved spectrum, $R(E,
f)$,  of a stationary process is defined as $\sqrt{P(E,f) df}$ (where $P(E,f)$
is the power spectral density function of the process).  Suppose now that the
X--ray  energy spectrum, $F(E,t)$,  of an AGN is given by

\[
F(E,t) = A(E)[B(E,t)+C(E)],
\]
where $B(E,t)$ represents the continuum, time variable emission of the source, 
$C(E)$ represents any spectral components that may be present and  do
not vary with time (i.e. X--ray reflection from distant material) and $A(E)$
stands for spectral components that modify the continuum emission (like cold
and/or warm absorption, absorption edges etc). 

In order to calculate the FRS  we need first to estimate the PSD of 
$F(E,t)$.  For a real valued, stationary process we have,

\[
P(E,f) = \int^{+\infty}_{-\infty} {\rm cos}(2\pi f\tau) ACF(E,\tau) d\tau,
\]

where 
\[
\begin{array}{ll}
ACF(E,\tau)=&\\
\\
\langle [F(E,t)- \langle F(E,t)\rangle] [F(E,t+\tau)- \langle 
F(E,t)\rangle] \rangle \nonumber & \end{array} 
\]
is the autocovariance function of $F(E,t)$ at lag $\tau$. 

For example,  let us consider the simple case of power-law like continuum,
where only the normalization (at 1 keV) varies with time, i.e.
$B(E,t)=K(t)E^{-\alpha}$. In this case,
$$\langle F(E,t)\rangle=A(E)[\langle K(t)\rangle E^{-\alpha}+C(E)]~.$$ 
It is easy to show that
$$ACF(E,\tau)=A^{2}(E) E^{-2\alpha}ACF_{K}(\tau),$$ 
where $ACF_{K}(\tau)$ is the
autocovariance function of $K(t)$. Consequently, 
$$P(E,f)=A^{2}(E) E^{-2\alpha}P_{K}(f),$$ 
and hence 
$$R(E, f)=A(E)E^{-\alpha} \sqrt{P_{K}(f)df}.$$
In other words, in the case when the X--ray energy spectrum of an AGN has a
power-law like shape with a slope of $\alpha$ and  varies only in
normalization, then the FR spectra also have a power-law shape, with slope
equal to $\alpha$, at {\it all} frequencies. Any constant components that may
be present in the time-average spectrum, and may complicate the correct
determination of the continuum slope, will not appear in the FR spectra. In
this case, it may be easier to determine the intrinsic spectral slope from the
model fitting of the FR spectra rather than of the time-average spectrum. 

We note that if $A(E)={\rm exp}[-n_{\rm H}\sigma(E)]$, i.e. the spectrum is
absorbed by an external absorber with constant column density, or $A(E)={\rm
exp}[-D(E/E_c)^{-3}]$ where $E_{c}$ is the  threshold energy of an absorption
edge (with constant absorption depth $D$),  then the FRS  will also show the 
signs of the absorption effects at the same energies, and with  the same
strength, as in the time-average spectrum of the source. 

Suppose now that $C(E)$ is also variable with time. For example, let us assume
that there is an emission line in the spectrum (at $E_{L}$)  whose flux varies
with time, i.e.  $$C(E,t)= \frac{L(t)}{\sqrt{\sigma
(2\pi)}} e^{-(E-E_L)^{2}/2\sigma^2}.$$ In this case, 
%
$$\langle F(E,t) \rangle = 
\nonumber $$
$$A(E)\left\{\langle K(t)\rangle E^{-\alpha}+
\frac{\langle L(t)\rangle}{\sqrt{\sigma
(2\pi)}} e^{-(E-E_L)^{2}/2\sigma^2}\right\}$$
%
where $\sigma$ is the width of the line.  In the simple case when the line is
produced by isotropic irradiation of matter in the innermost part of an
accretion disk by the X--ray continuum source, one would expect the bulk of the
line to respond  in synchrony with the continuum variations and hence 
$L(t)\sim cK(t)$  (where $c$ is  a constant). One can show that in this case 
$$ACF(E,\tau)=A(E)^{2} \times$$
$$\times \left\{(E)^{-\alpha}+ \frac{c}{\sqrt{\sigma
(2\pi)}}e^{-(E-E_L)^{2}/2\sigma^2}\right\}^2 ACF_{K}$$ 
and hence 
$$P(E,f)=A(E)^{2} \times$$ 
$$\left\{(E)^{-\alpha}+ \frac{c}{\sqrt{\sigma
(2\pi)}}e^{-(E-E_L)^{2}/2\sigma^2} \right\}^2 P_{K}(f).$$ 
Consequently, the FRS  will
be given by 
$$R(E,f)= A(E) \times$$ 
$$\times \left\{ E^{-\alpha} + \frac{c}{\sqrt{\sigma (2\pi)}}
e^{-(E-E_L)^{2}/\sigma^2}\right\}\sqrt{P_{K}(f)df} ~.$$
The amplitude of the line relative to the continuum (i.e. the line's EW) will
be the same in both the time averge and the FR spectra of all frequencies. 

Finally let us consider the case of a power law with a variable slope, i.e. the
case when $B(E,t)=K(E/E_P)^{-\alpha(t)}$ (where $E_P$ is the pivot energy). If
$\sigma^{2}_{\alpha}$ is significantly smaller than $\alpha$ (a condition that
probably holds in the case of many AGN where typically $\alpha\sim 1$ and 
$\sigma^{2}_{\alpha}\sim 0.1)$, then it can be shown that 
$$ACF(E,\tau)=A(E)^{2} \times$$
$$ \times
K^{2} (E/E_{P})^{-2\langle \alpha \rangle}{\rm ln}^{2}(E/E_{C})ACF_{\alpha}(\tau).$$ 
Subsequently, 
$$P(E,f)= A(E)^{2} K^{2} (E/E_{P})^{-2\langle \alpha \rangle}
{\rm ln}^2(E/E_P)P_{\alpha}(f),$$ 
and the FR spectra will be given by the relation, 
$$R(E,f)= A(E)K\left(\frac{E}{E_{P}}\right)^{-\langle \alpha \rangle}
{\rm ln}\left(\frac{E}{E_{P}}\right)\sqrt{P_{\alpha}(f)df}.$$ 

If $E_{P}$ is located within the energy range ($E_{\rm min}-E_{\rm max}$) at
which the FR spectra are estimated, the Fourier-resolved spectra will {\it not}
have a power-law shape. Instead, significant curvature will appear around
$E_{P}$. If on the other hand $E_{P}$ is outside the energy range that is
considered, the deviation of the Fourier-resolved spectra from a power law
shape with slope equal to the time-average $\alpha$ ($\langle \alpha \rangle$)
may not be noticed. Hence, even in the case of a pivoting energy spectrum, we
expect the FRS to have the {\it same} slope  ($\sim \langle \alpha \rangle$) at
all frequencies.




\clearpage
\begin{deluxetable}{lcrcccc}
\tablecolumns{5}
\tablecaption{Details of the observations used in the present work.}
\tablehead{
\colhead{Name} & \colhead{Date} & \colhead{Exp. Time} & \colhead{Obs ID}
& \colhead{PI} & \colhead{Filter} & \colhead{Mode}}
\startdata
Mrk~766  & 2001/05/20 & 100 ks & 0109141301 & Mason     & Medium  & Small Window\\
NGC~3516 & 2001/04/10 &  88 ks & 0107460601 & Mushotzky & Thin    & Small Window\\
NGC~3783 & 2001/12/17 & 125 ks & 0112210201 & Kaastra   & Medium  & Small Window\\
         & 2001/12/19 & 123 ks & 0112210501 & Kaastra   & Medium  & Small Window\\
NGC~4051 & 2001/05/16 & 101 ks & 0109141401 & Mason     & Medium  & Small Window\\
         & 2002/11/22 &  34 ks & 0157560101 & Jansen    & Medium  & Large Window\\
\ark     & 2005/01/05 & 100 ks & 0206400101 & Papadakis & Medium  & Small Window\\
\enddata
\label{tab:obs}
\end{deluxetable}

\clearpage
\begin{deluxetable}{cccc}
\tablecolumns{4}
\tablecaption{The \mrk\ best model fitting results
(N$_{\rm H}=1.7\times 10^{20} {\rm cm}^{-2}$)}
\tablehead{
\colhead{Parameter} & \colhead{Mean} & \colhead{LF} & \colhead{HF} }
\startdata

${\Gamma}$                      &$2.14\pm 0.02$
&  $2.19{\pm}0.04$            & $1.93\pm 0.11$    \\
E$_{\rm line}$ (keV)            &$6.45\pm 0.10$
&  $6.45$\tablenotemark{a}             &  $-$              \\
$\sigma_{\rm line}$ (eV)        &$480\pm100$
&  $100$\tablenotemark{a}              &  $-$              \\
EW (eV)                         &$230^{+20}_{-50}$
&  $60^{+70}_{-35}$            &  $-$              \\
${\chi}^{2}/$d.o.f.             & 6.2/5
& 5.5/7                       & 12.2/7             \\

\enddata
\label{tab:766specres}
\tablenotetext{a}{Parameter kept fixed}
\end{deluxetable}

\clearpage
\begin{deluxetable}{cccc}
\tablecolumns{4}
\tablecaption{The NGC~3516 best model fitting results
(N$_{\rm H}=2.9\times 10^{20} {\rm cm}^{-2}$)}
\tablehead{
\colhead{Parameter} & \colhead{Mean} & \colhead{LF} }
\startdata

${\Gamma}$ (keV)                            &$1.34\pm 0.03$  & $1.34\pm 0.11$ \\
E$_{\rm line}$ (keV)                        &$6.4$\tablenotemark{a}   & $-$     \\
$\sigma_{\rm line}$ (eV)                    &$100$\tablenotemark{a}   & $-$     \\
EW (eV)                             &$130\pm 0.15$   & $-$     \\
${\chi}^{2}/$d.o.f.                         &7.7/4           &10.6/8   \\

\enddata
\label{tab:3516specres}
\tablenotetext{a}{Parameter kept fixed}
\end{deluxetable}

\clearpage
\begin{deluxetable}{ccc}
\tablecolumns{4}
\tablecaption{The NGC~3783 best model fitting results
(N$_{\rm H}=8.3\times 10^{20} {\rm cm}^{-2}$)}
\tablehead{
\colhead{Parameter} & \colhead{Mean} & \colhead{LF} }
\startdata

${\Gamma}$                      &$1.46\pm 0.01$
&  $1.94{\pm}0.02$            \\
E$_{\rm line}$ (keV)            &$6.4$\tablenotemark{a}
&  $6.05\pm0.31$             \\
$\sigma_{\rm line}$ (eV)        &$100$\tablenotemark{a}
&  $100$\tablenotemark{a}              \\
EW (eV)                         &$100\pm 16$
&  $200^{+69}_{-73}$            \\
E$_{\rm edge}$ (keV)         & $7.1$\tablenotemark{a}
& $-$                          \\
$\tau_{\rm edge}$            & $0.12\pm0.07$
& $-$                         \\
${\chi}^{2}/$d.o.f.             & 5.3/6
& 7.8/6                       \\

\enddata
\label{tab:3783specres}
\tablenotetext{a}{Parameter kept fixed}
\end{deluxetable}

\clearpage
\begin{deluxetable}{cccccc}
\tablecolumns{6}
\tablecaption{The 2001 and 2002 NGC~4051 best model fitting results
(N$_{\rm H}=1.3\times 10^{20} {\rm cm}^{-2}$)}
\tablehead{
\colhead{Parameter} & \colhead{Mean(01)} & \colhead{LF(01)} & \colhead{HF(01)}
& \colhead{Mean(02)} &  \colhead{LF(02)}  }
\startdata

${\Gamma}$                      &$1.79\pm 0.02$  & $2.17\pm 0.04$
&  $1.99{\pm}0.19$            & $0.88\pm 0.03$   & $1.83\pm 0.17$  \\
E$_{\rm line}$ (keV)            &$6.10\pm 0.20$  &   $6.10$\tablenotemark{a}
&  $-$             &  $6.4$\tablenotemark{a}   & $-$          \\
$\sigma_{\rm line}$ (eV)        &$100$\tablenotemark{a}   &  $100$\tablenotemark{a}
&  $-$              &  $100^{\rm a}$  & $-$           \\
EW (eV)                         &$84\pm 12$  &  $41^{+38}_{-35}$
&  $-$            &  $191\pm 20$ & $-$              \\
E$_{\rm edge}$ (keV)         & $7.1$\tablenotemark{a}  & $-$
& $-$                         & $7.5\pm 0.1$    &$-$             \\
$\tau_{\rm edge}$            & $0.11\pm 0.03$     & $-$
& $-$                         & $0.75\pm 0.12$    & $-$            \\
${\chi}^{2}/$d.o.f.             & 5.1/5          & $12.8/7$
& 7.9/8                       & 9.8/5 & 6.5/10              \\

\enddata
\label{tab:4051specres}
\tablenotetext{a}{Parameter kept fixed}
\end{deluxetable}

\clearpage
\begin{deluxetable}{cccc}
\tablecolumns{4}
\tablecaption{The \ark\ best model fitting results
(N$_{\rm H}=6.4\times 10^{20} {\rm cm}^{-2}$)}
\tablehead{
\colhead{Parameter} & \colhead{Mean} & \colhead{LF} & \colhead{HF} }
\startdata

${\Gamma}$                      &$2.45\pm 0.02$
&  $2.51{\pm}0.05$            & $2.53\pm 0.46$    \\
E$_{\rm line}$ (keV)            &$6.65\pm 0.20$
&  $-$             &  $-$              \\
$\sigma_{\rm line}$ (eV)        &$100$\tablenotemark{a}
&  $-$              &  $-$              \\
EW (eV)                         &$84^{+19}_{-17}$
&  $-$            &  $-$              \\
${\chi}^{2}/$d.o.f.             & 3.8/6
& 9.6/8                       & 6.1/7             \\

\enddata
\label{tab:arkspecres}
\tablenotetext{a}{Parameter kept fixed}
\end{deluxetable}

\clearpage


\figcaption[fig1.eps]{
The full band, $0.2 - 10$ keV, light curves of \mrk, NGC~3516, and NGC~3783 in
bins of size 100 s. The light curves are normalized to the lowest count rate,
in order to reveal clearly the observed variability amplitude in each case.
\label{fig:lc1}}

\figcaption[fig2.eps]{
Same as with Figure~\ref{fig:lc1} in the case of the NGC~4051 and \ark\ full
band light curves.
\label{fig:lc2}}

\figcaption[fig3.eps]{
Upper panel: The time-average (\ie, ``mean"), LF and HF Fourier-resolved
spectra of \mrk\ (open circles, filled squares and triangles, respectively).
The dashed lines show the best fitting model in the case of the time-average
spectrum, and the best fitting simple Power-Law model in the case of the LF and
HF spectra.  Lower panels: Plot of the best model fitting residuals in the case
of the time-average spectrum, and of the best PL model fitting residuals in the
case of the LF and HF spectra.
\label{fig:mrk}}

\figcaption[fig4.eps]{
Same as with Figure~\ref{fig:mrk} in the case of the time-average and the
LF spectrum of NGC~3516.
\label{fig:3516}}

\figcaption[fig5.eps]{
Same as with Figure~\ref{fig:mrk} in the case of the time-average and the
LF spectrum of NGC~3783.
\label{fig:3783}}

\figcaption[fig6.eps]{
Same as with Figure~\ref{fig:mrk} in the case of the time-average, the
LF and HF spectrum of NGC~4051 (May 2001 observation).
\label{fig:4051-1}}

\figcaption[fig7.eps]{
Same as with Figure~\ref{fig:mrk} in the case of the time-average and the
LF spectrum of NGC~4051 (November 2002 observation).
\label{fig:4051-2}}

\figcaption[fig8.eps]{
Same as with Figure~\ref{fig:mrk} in the case of the time-average , the
LF and HF spectrum of \ark.
\label{fig:ark}}

\figcaption[fig9.eps]{
The Fourier-resolved spectral slope of AGN, including the results of Papadakis
\etal\ (2005) on MCG -6-30-15, plotted as a function of the Fourier frequency
normalized to the Keplerian frequency at 3R$_{\rm S}$ (filled squares). The
\ark\ and NGC~3873 results are plotted with different symbols for the reasons
discussed in the text. The open squares and circles show the Cyg X-1 results in
the case when the system is at its Low and High state, respectively. The solid
line shows the best fitting model to the AGN data. The long dashed and dotted
lines have the same slope but are shifted to lower and higher  $\Gamma_{\rm
FRS}$ values, respectively, in order to match the Cyg X-1 results in LS and HS.
\label{fig:gamma}}

\figcaption[fig10.eps]{
The EW of the iron-line which is detected in the LF spectra of NGC~4051, \mrk,
and NGC~3783 plotted as a function of the Fourier frequency normalized to the
Keplerian frequency at 3R$_{\rm S}$ (filled squares; the sequence as given 
above from left to right). Open squares and circles
show the Cyg X-1 results in the case when the system is at its Low and High
states, respectively. The dashed line shows the best fitting model to the Cyg
X-1 at frequencies above $\sim 0.005$ of $f_{\rm K}(3\rm R_{S})$.
\label{fig:ew}}

\clearpage

\begin{figure}
\epsscale{.7}
\figurenum{\ref{fig:lc1}}
\plotone{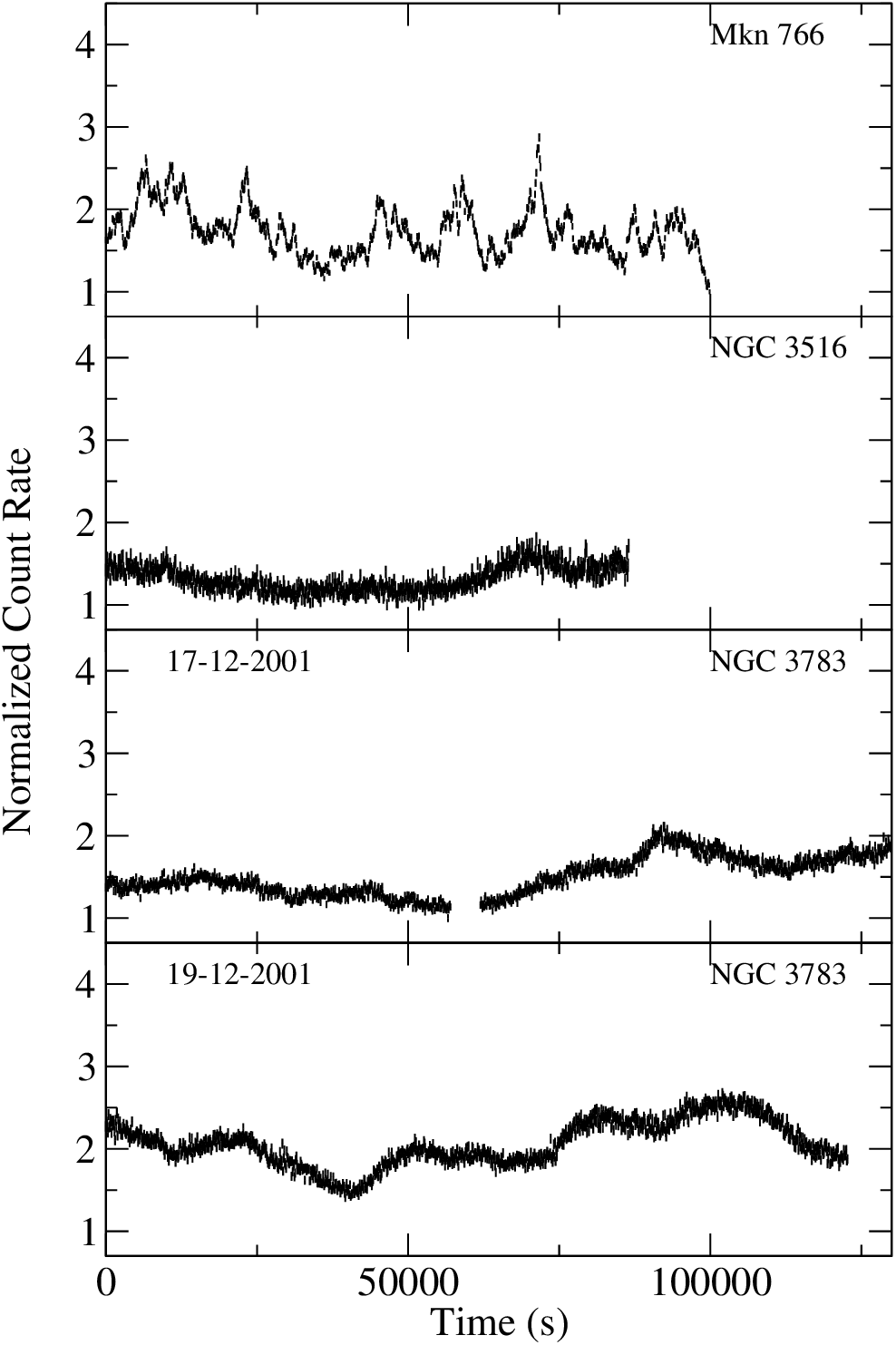}
\end{figure}

\begin{figure}
\figurenum{\ref{fig:lc2}}
\plotone{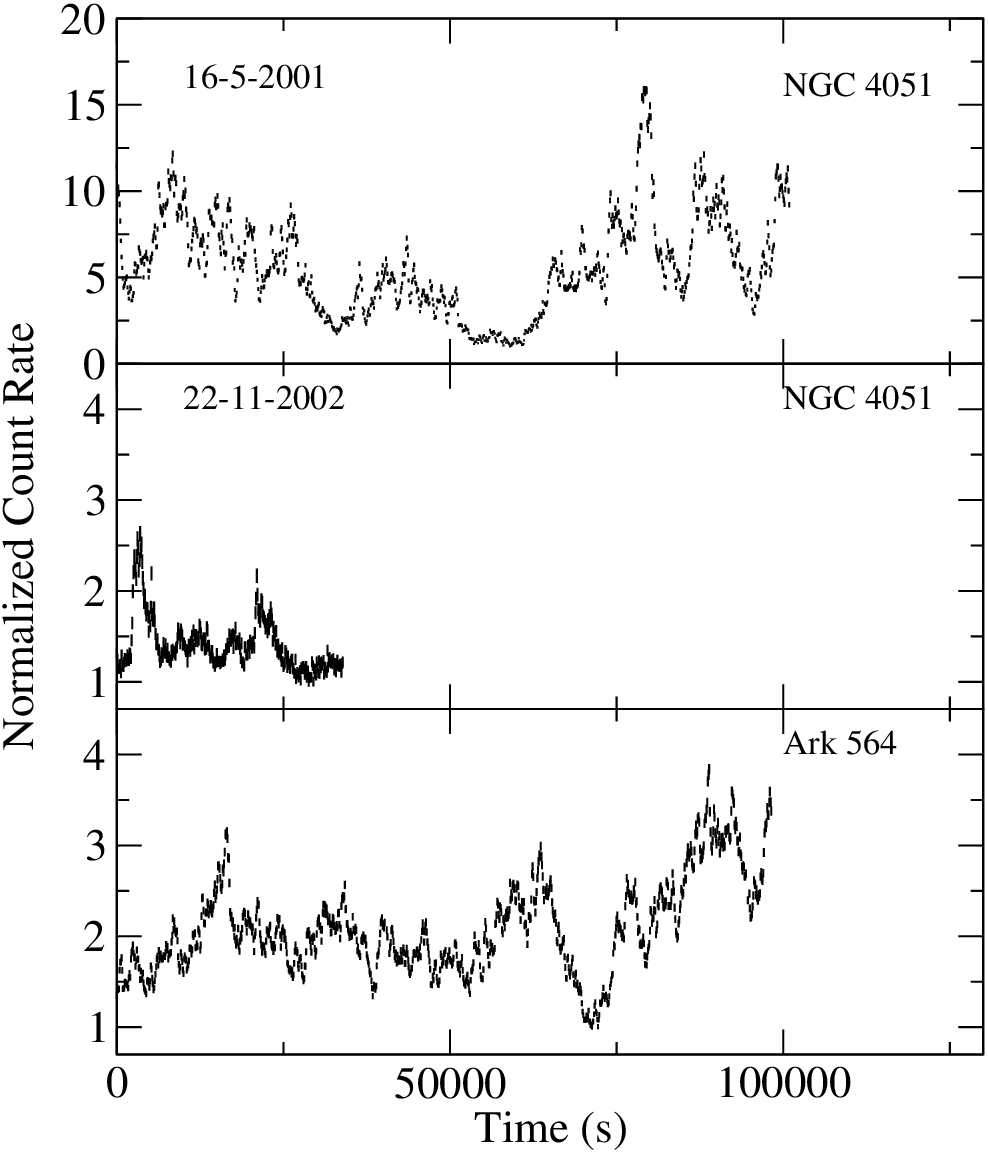}
\end{figure}

\begin{figure}
\figurenum{\ref{fig:mrk}}
\plotone{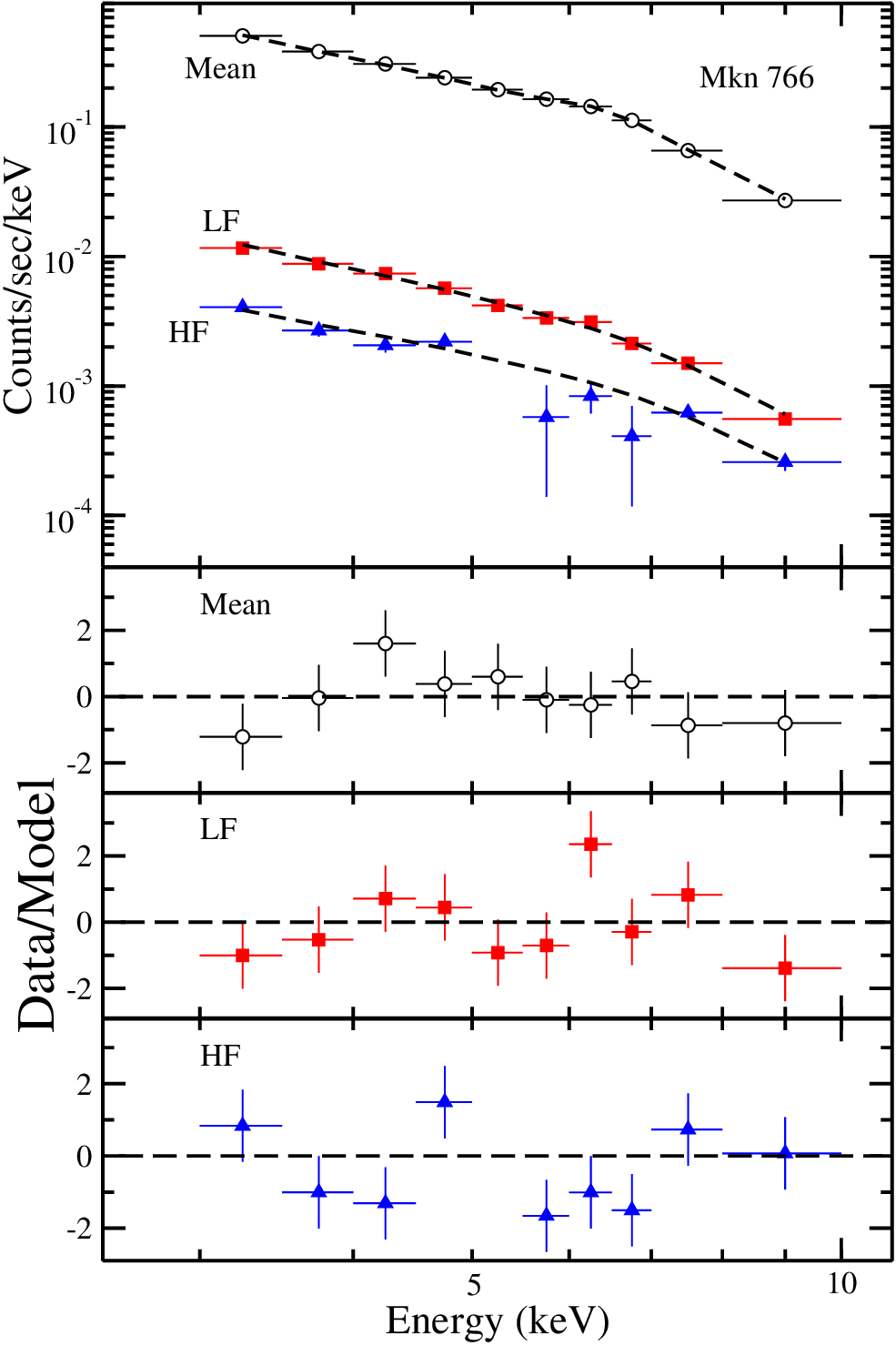}
\end{figure}

\begin{figure}
\figurenum{\ref{fig:3516}}
\plotone{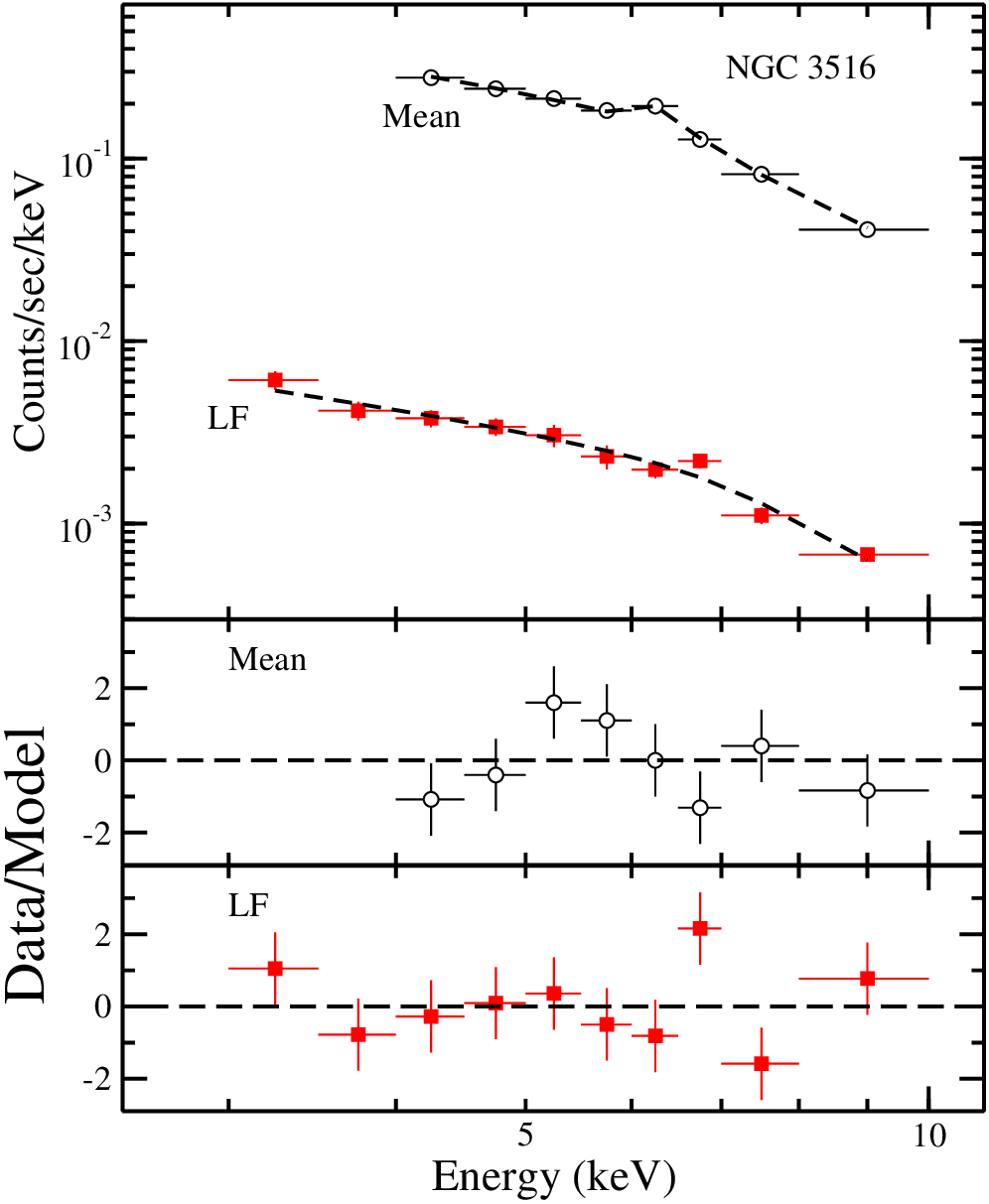}
\end{figure}

\begin{figure}
\figurenum{\ref{fig:3783}}
\plotone{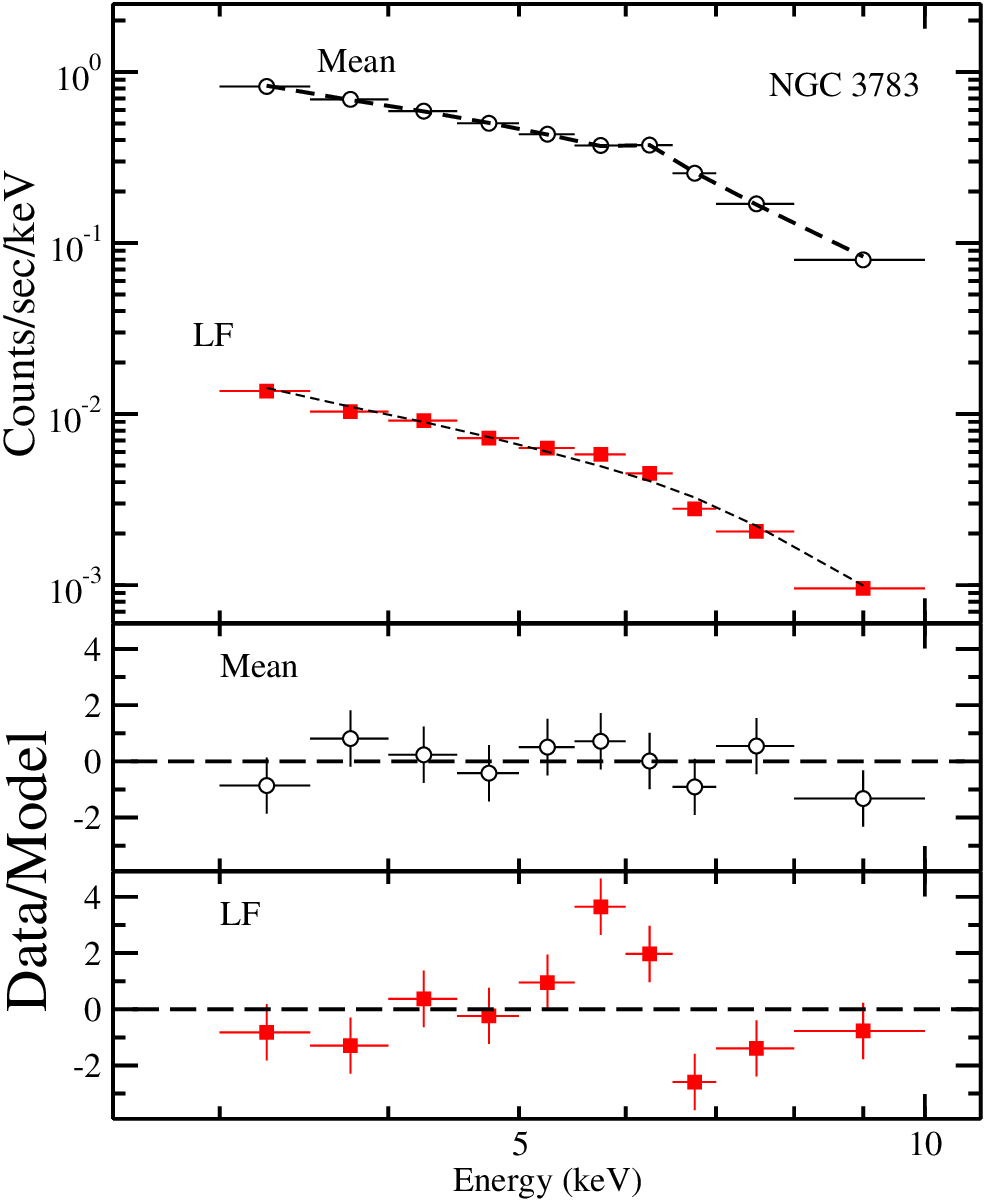}
\end{figure}

\begin{figure}
\figurenum{\ref{fig:4051-1}}
\plotone{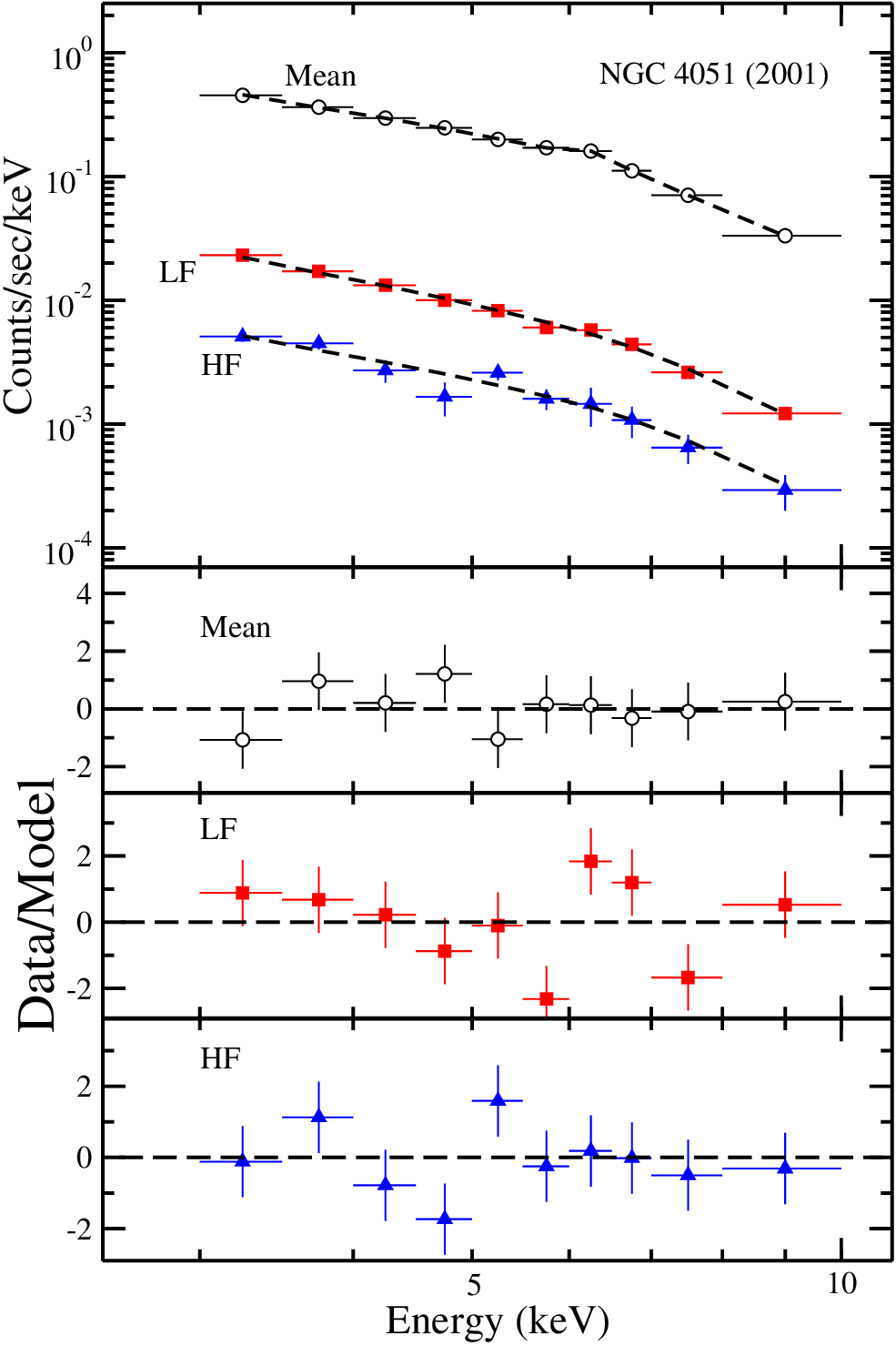}
\end{figure}

\begin{figure}
\figurenum{\ref{fig:4051-2}}
\plotone{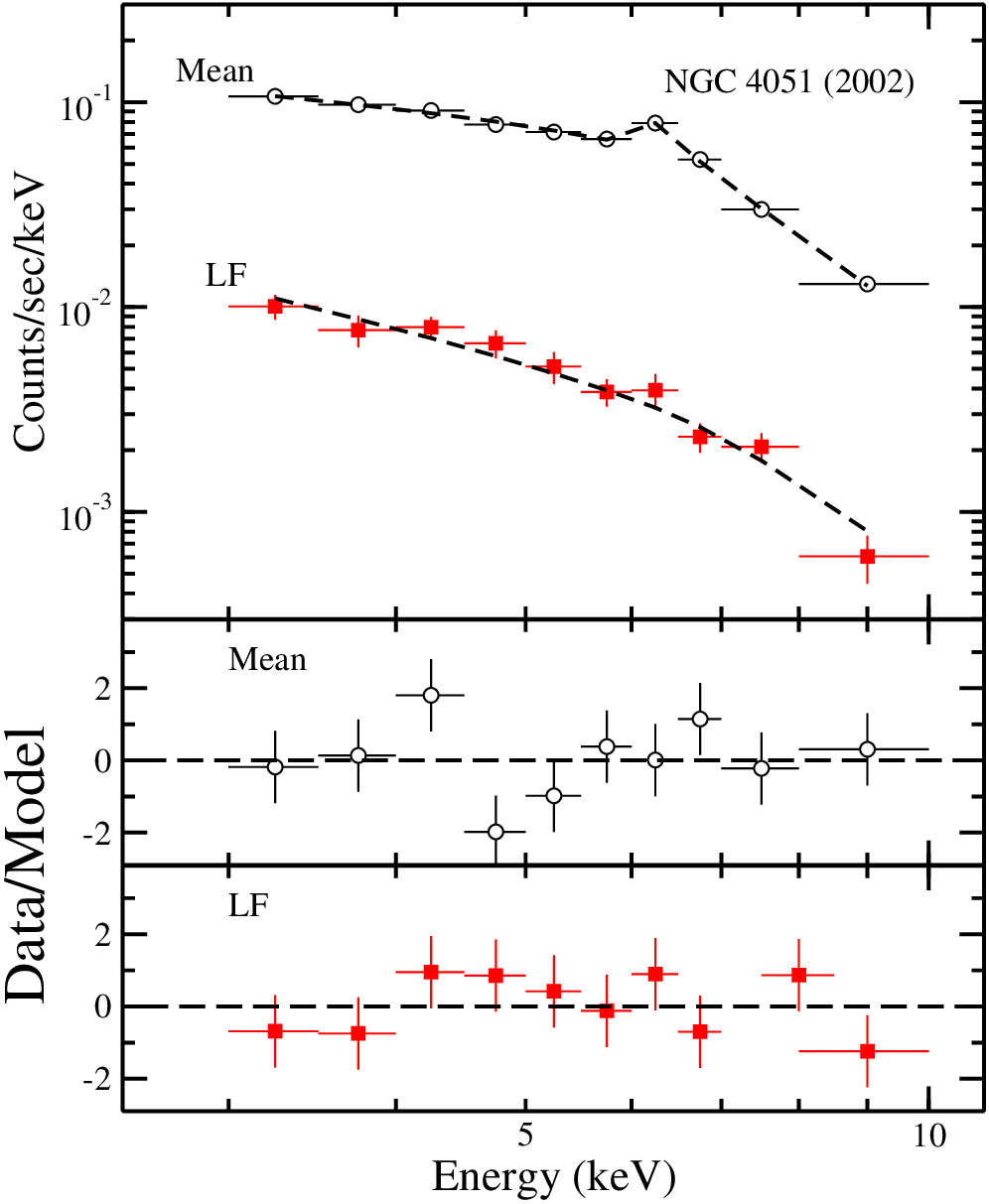}
\end{figure}

\begin{figure}
\figurenum{\ref{fig:ark}}
\plotone{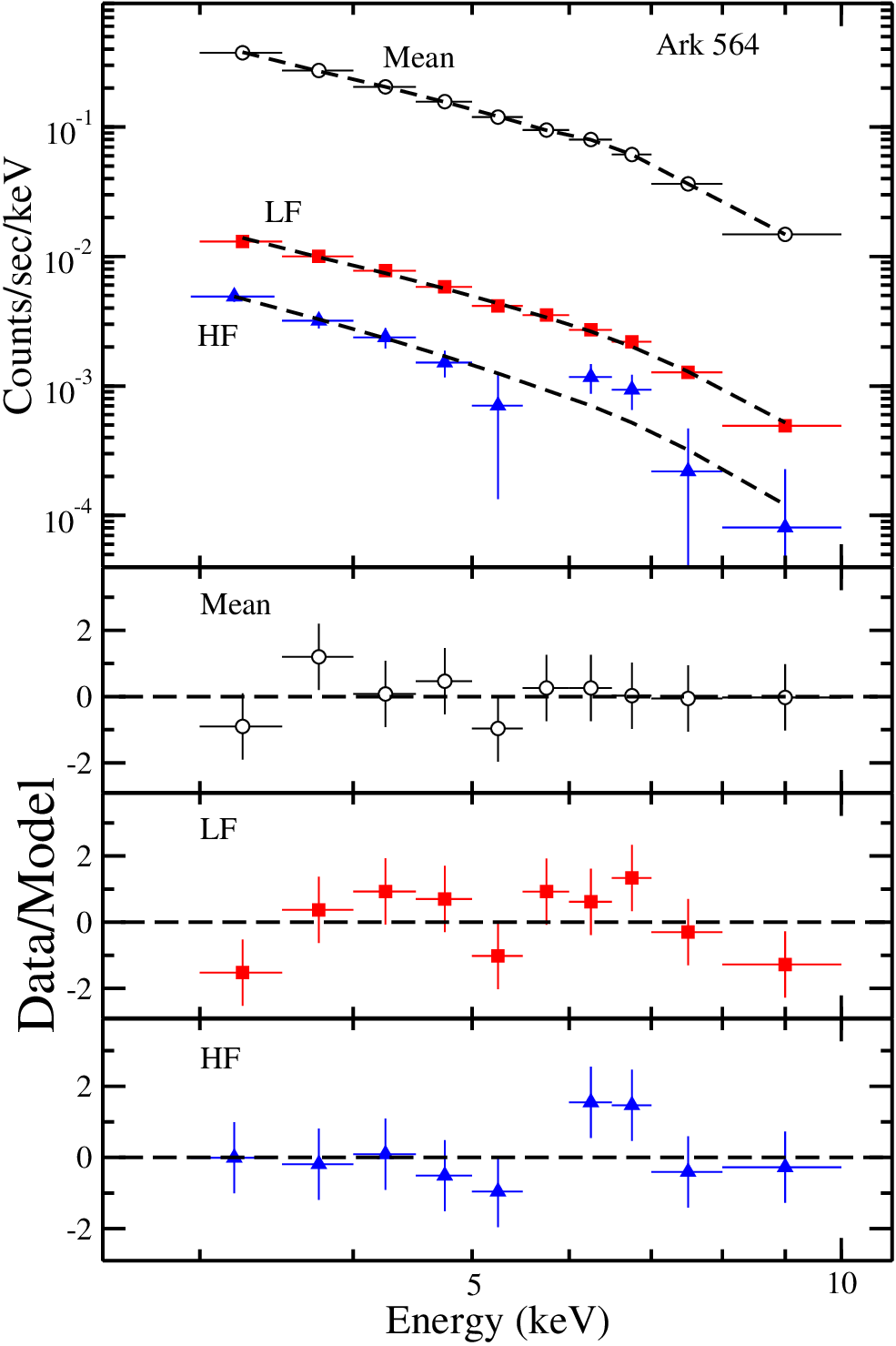}
\end{figure}

\begin{figure}
\figurenum{\ref{fig:gamma}}
\plotone{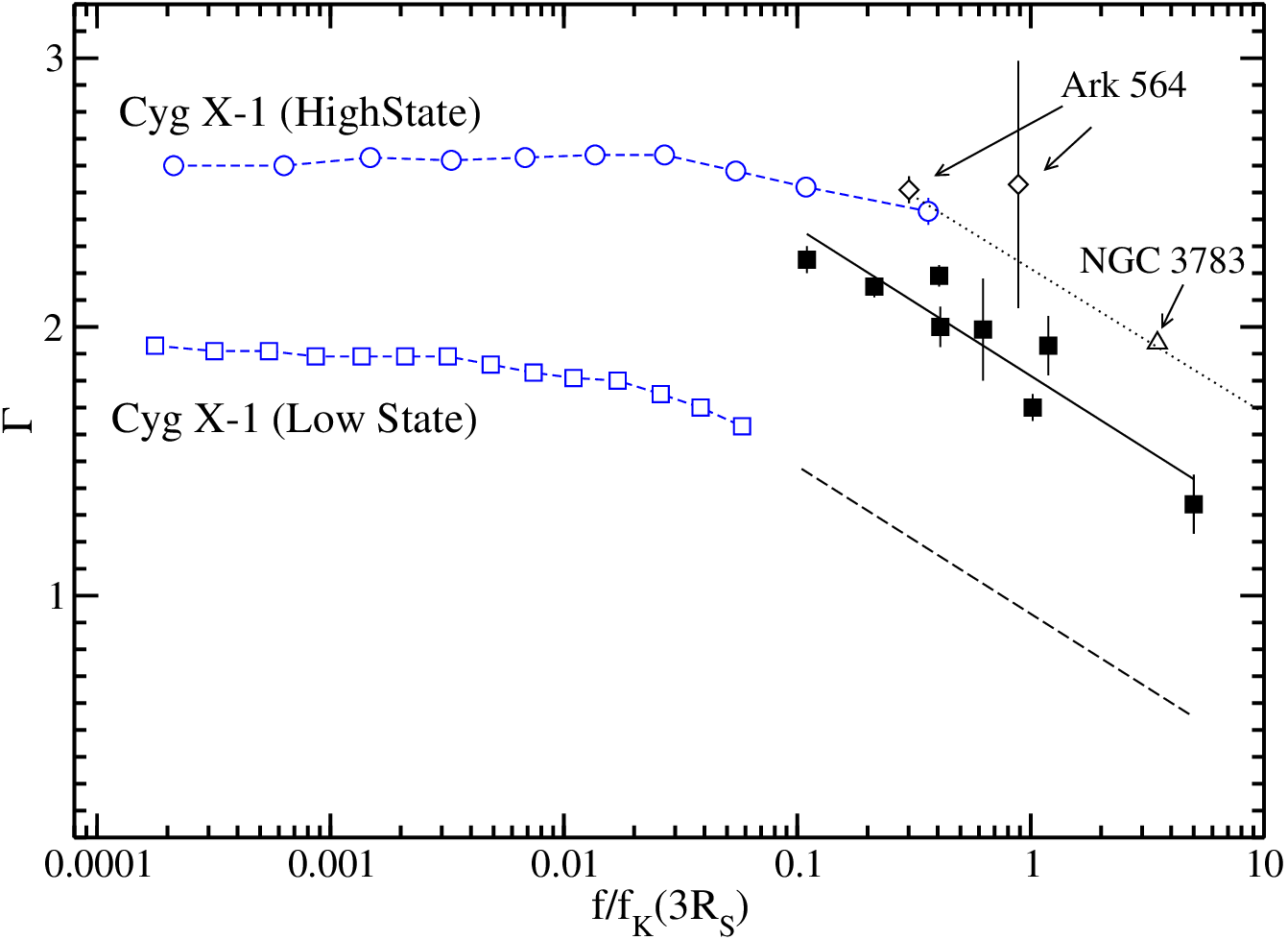}
\end{figure}

\clearpage
\begin{figure}
\figurenum{\ref{fig:ew}}
\plotone{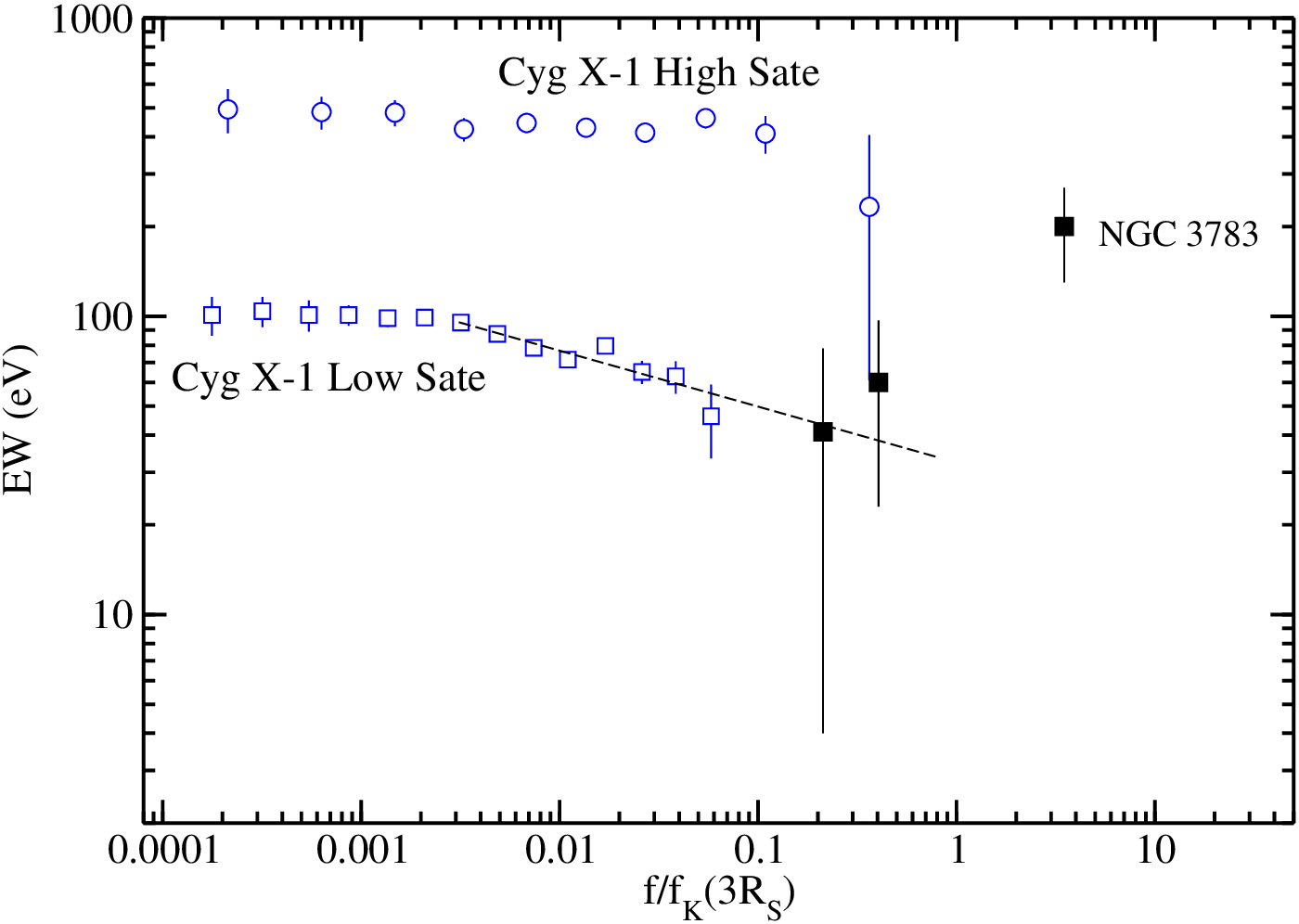}
\end{figure}


\begin{references}

\reference{}Arevalo, P., Papadakis, I.E., Uttley, P., M$^{\rm c}$Hardy, I.M., \& Brinkmann,
W. 2006, MNRAS, 372, 401

\reference{}Botte, V., Ciroi, S., Rafanelli, P., \& Di Mille, F. 2004, AJ, 127, 3168

\reference{}Dickey, J.M., \&  Lockman, F.J. 1990, ARA\&A, 28, 215

\reference{}Edelson, R., Turner, T.J., Pounds, K., Vaughan, S., Markowitz, A., 
Marshall, H., Dobbie, P., \& Warwick, R. 2002, ApJ, 568, 610

\reference{}Gilfanov, M., Revnivtsev, M., \& Molkov, S. 2003, A\&A, 410, 217

\reference{}Gilfanov, M., \& Revnivtsev, M. 2005, AN, 326, 812

\reference{}Iwasawa, K., et al., 1996, MNRAS, 282, 1038

\reference{}Iwasawa, K., Fabian, A.C., Young, A.J., Inoue, H. \& Matsumoto, C.
1999, MNRAS, 306, L19

\reference{}Kazanas, D., Hua, X.-M., \& Titarchuk, L. 1997, ApJ, 480, 735

\reference{}Lee, J.C., Fabian, A.C., Reynolds, C.S., Brandt, W.N. \& Iwasawa, K. 2000,
MNRAS, 318, 857

\reference{}Markowitz, A., Edelson, R. \& Vaughan, S. 2004, ApJ, 598, 935

\reference{}Miller, L., Turner, T.J., Reeves, J.N., George, I.M., Porquet, D., Nandra,
K., \& Dovciak, M. 2006, A\& A, 453, L13

\reference{}M$^{\rm c}$Hardy, I.M., Gunn, K.F., Uttley, P., \& Goad, M. R. 2005, MNRAS, 359,
1469

\reference{}Nandra, K., George, I. M., Mushotzky, R. F., Turner, T. J., \& Yaqoob, T.
1997, ApJ, 477, 602

\reference{}Narayan, R., \& Yi, I. 1994, ApJ, 428, L13

\reference{}Nayakshin, S., Kazanas, D., \& Kallman, T. R. 2000, ApJ, 537, 833

\reference{}Papadakis, I.E., Brinkmann, W., Page, M.J., M$^{\rm c}$Hardy, I., \& Uttley, U. 2007,
A\&A, 461, 931
 
\reference{}Papadakis, I.E., Kazanas, Demosthenes, \& Akylas, A. 2005, ApJ, 2005, 631, 727

\reference{}Peterson, B. M., et al. 2004, ApJ, 613, 682

\reference{}Pounds, K.A., Reeves, J.N., Page, K.L.; Wynn, G.A., \& O'Brien, P.T. 2003, 342,
1147

\reference{}Pounds, K.A., Reeves, J.N., King, A.R., \& Page, K.L. 2004, MNRAS, 350, 10

\reference{}Poutanen, J., \& Fabian, A. C. 1999, MNRAS, 306, L31

\reference{}Priestley, M.B. 1989, Spectral Analysis and Time Series (London, Academic Press
Limited)

\reference{}Reeves, J.N., Nandra, K., George, I.M., Pounds, K.A., Turner, T.J.,  \& Yaqoob,
T. 2004, ApJ, 602, 648

\reference{}Reig, P., Papadakis, I.E., Shrader, C.R., \& Kazanas, D. 2006, ApJ, 644, 424

\reference{}Revnivtsev, M., Gilfanov, M., \& Churazov, E. 1999, A\&A, 347, 23

\reference{}Revnivtsev, M., Gilfanov, M., \& Churazov, E. 2000, MNRAS, 316, 923

\reference{}Revnivtsev, M., Gilfanov, M., \& Churazov, E. 2001, A\&A, 380, 520

\reference{}Sobolewska, M.A., \& $\dot{\rm Z}$ycki, P.T. 2006, MNRAS, 370, 405

\reference{}Taylor, R.D., Uttley, P., \& M$^{\rm c}$Hardy, I. M. 2003, MNRAS, 342, L31

\reference{}Turner, T.J., Kraemer, S.B., George, I.M., Reeves, J.N., \& Bottorff, M.C. 2005,
ApJ, 618, 155

\reference{}Uttley, P., Taylor, R.D., M$^{\rm c}$Hardy, I.M., Page, M.J., Mason, K.O., Lamer, G.,
\& Fruscione, A. 2004, 347, 1345

\reference{}Vaughan, S. \& Edelson, R. 2001, ApJ, 548, 694

\reference{}Vaughan, S. \& Fabian, A.C. 2004, MNRAS, 348, 1415

\reference{}Woo, J-H, \& Urry, C.M. 2002, ApJ, 579, 530

\reference{}Zdziarski, A. A., Lubinski, Piotr \& Smith, D. A. 1999, MNRAS, 303,
L11

\reference{}$\dot{\rm Z}$ycki, P. 2003, MNRAS, 340, 639 

\reference{}$\dot{\rm Z}$ycki P. 2004, MNRAS, 351, 1180

\end{references}
\end{document}